\documentclass[11pt,a4paper]{article}
\pdfoutput=1
\usepackage{jcappub}
\usepackage{ifthen}
\usepackage[T1]{fontenc}
\usepackage{lmodern}
\usepackage[utf8]{inputenc}
\usepackage[english]{babel}
\usepackage[usenames,dvipsnames,svgnames]{xcolor}
\usepackage{amsmath,amssymb}
\usepackage[normalem]{ulem}  % To remove in the final version
\usepackage{array,multirow,graphicx}
\graphicspath{{figures/}}
\usepackage{float}
\usepackage{subfig}
\usepackage{comment}

\usepackage{cancel}
\usepackage[colorinlistoftodos]{todonotes}
\usepackage{import}

\usepackage{natbib}

\newcommand{\dd}{\textrm{d}}

\title {\huge The CMB cold spot under the lens: \\
{\huge ruling out a supervoid interpretation}}

\author[a,*]{Stephen Owusu,}
\author[b,*]{Pedro da Silveira Ferreira,}
\author[c,d]{Alessio Notari,}
\author[a,b,e,f]{and Miguel Quartin}

\affiliation[a]{Instituto de Física, Universidade Federal do Rio de Janeiro, 21941-972, Rio de Janeiro, RJ, Brazil}
\affiliation[b]{Observatório do Valongo, Universidade Federal do Rio de Janeiro, 20080-090, Rio de Janeiro, RJ, Brazil}
\affiliation[c]{Departament de F\'isica Fondamental i Institut de Ci\'encies del Cosmos, Universitat de Barcelona, Mart\'i i Franqu\'es 1, E-08028 Barcelona, Spain}
\affiliation[d]{Galileo Galilei Institute for theoretical physics, Centro Nazionale INFN di Studi Avanzati
Largo Enrico Fermi 2, I-50125, Firenze, Italy}
\affiliation[e]{Institute of Theoretical Physics, Heidelberg University, Philosophenweg 16, 69120 Heidelberg, Germany}
\affiliation[f]{PPGCosmo, Universidade Federal do Espírito Santo, 29075-910, Vitória, ES, Brazil}
\affiliation[*]{{\bf these authors contributed equally to this work}}

\emailAdd{mquartin@if.ufrj.br}

\abstract{The Cosmic Microwave Background (CMB) anisotropies are thought to be statistically isotropic and Gaussian. However, several anomalies are observed, including the CMB Cold Spot, an unexpected cold $\sim\!10^{\circ}$ region with $p\,$-value $\lesssim 0.01$ in standard $\Lambda$CDM. One of the proposed origins of the Cold Spot is an unusually large void on the line of sight, that would generate a cold region through the combination of integrated Sachs-Wolfe and Rees-Sciama effects.  In the past decade extensive searches were conducted in large scale structure surveys, both in optical and infrared, in the same area for $z\lesssim 1$ and did find evidence of large voids, but of depth and size able to account for only a fraction of the anomaly. Here we analyze the lensing signal in the Planck CMB data and rule out the hypothesis that the Cold Spot could be due to a large void located anywhere between us and the surface of last scattering. In particular, computing the evidence ratio we find that a model with a large void is disfavored compared to $\Lambda$CDM, with odds 1 : 13 (1 : 20) for SMICA (NILC) maps, compared to the original odds 56 : 1 (21 : 1) using temperature data alone. }


\begin{document}

\notoc
\maketitle

%%%%%%%%%%%%%%%%%%%%%%%%%%%%%%%%%%%%%%%%
%%%%%%%%%%%%%%%%%%%%%%%%%%%%%%%%%%%%%%%%
\section{Introduction} \label{sec:intro}
%%%%%%%%%%%%%%%%%%%%%%%%%%%%%%%%%%%%%%%%
%%%%%%%%%%%%%%%%%%%%%%%%%%%%%%%%%%%%%%%%

The Cosmic Microwave Background (CMB) has become one of the most useful tools to understand the evolution of the universe, leading to the establishment of the $\Lambda$CDM standard cosmological model. Primordial fluctuations (anisotropies) on the CMB are thought to be statistically isotropic and nearly Gaussian, consistently with predictions from primordial inflation. However, several so-called anomalies are often discussed in the literature, including: the quadrupole-octopole alignment \cite{deOliveira-Costa:2003utu,Copi:2010na,Notari:2015kla}, the hemispherical power asymmetry \cite{Eriksen:2003db,Bernui:2006ft,Hansen:2008ym,Quartin:2014yaa} and the case of interest, the anomalous Cold Spot (CS)~\cite{Vielva:2003et,Cruz:2004ce,Cruz:2006sv}.
The significance of these anomalies is often shown to be around the $3\sigma$ level, but the employed statistics are sometimes disputed, as they depend on \emph{a posteriori} selections of data subsets~\citep{Bennett:2010jb,Naidoo:2017woy}, and other statistics designed from the start to avoid such selections are instead consistent with standard $\Lambda$CDM~\citep{Oliveira:2018sef}. For more recent reviews on the CMB anomalies see~\citep{Schwarz:2015cma,Muir:2018hjv}.

The CS is a cold $\sim10^\circ$ radius region in the CMB temperature map, centered around the Galactic coordinates $(l,\,b)\sim(209^\circ, \,-57^\circ)$, with an unusual combination of angular size, coldness and surrounding hot ring. Since its detection in the WMAP data, through an analysis based on Spherical Mexican Hat Wavelets~\citep{Vielva:2003et}, it has been one of the most extensively studied large-scale CMB anomalies~\citep{Cruz:2004ce,Cruz:2008sb}. The non-Gaussianity and the morphology of the CS were studied in~\cite{Cruz:2006sv, Cruz:2004ce}, and the conclusion was that the probability to find such a region from standard primordial fluctuations in $\Lambda$CDM is less than 2\%.  Most recently the Planck Collaboration also confirmed the existence of the CS anomaly, estimating its significance at about 1\%, based on the $\Lambda$CDM model~\citep{Planck:2015igc}.

There are several proposals to account for its origin, beyond a simple statistical fluke. These include the presence of a large void~\cite{Inoue:2006rd, Inoue:2006fn, Gurzadyan:2014yxa} or a hypothetical cosmic texture~\cite{Cruz:2007pe,Das:2008es} somewhere along the line of sight or a cosmic bubble collision~\cite{Feeney:2010jj}.  Clearly a combination of any of these would also be possible.
Among these possibilities the presence of a void may seem the least exotic, but nevertheless such a void would also be anomalously large in the $\Lambda$CDM model. In this paper we will focus on this void hypothesis.

An intervening void imprints an extra negative temperature fluctuation due to a combination of the integrated Sachs-Wolfe (ISW) effect, which is associated to the linear evolution of the gravitational potential due to dark energy, and the Rees-Sciama (RS) effect, which is a second-order ISW effect~\cite{Rees:1968zza,Inoue:2011st}.

The first indication of anomalously large voids in the CS region came from
pencil beam surveys at redshifts $z<0.3$  \cite{Granett:2009aw,Bremer:2010jn}. This was reinforced later with the infrared WISE-2MASS galaxy catalogue, which detected a void aligned with the CS, with an estimated size of $\sim 200 h^{-1} $Mpc, an average density contrast of $\bar{\delta} \approx -0.15$, and centered at redshift of $z \approx 0.15$~\cite{Szapudi:2014zha,Kovacs:2015hew}. Recently, the Dark Energy Survey (DES) confirmed the presence of a void in the same region \cite{Kovacs:2021wnc}, which they refer to as the Eridanus supervoid. Nonetheless, the expected ISW and RS effects caused by the detected structures are not large enough~\cite{Nadathur:2014tfa,Finelli:2014yha,Kovacs:2021wnc} to explain the temperature profile associated with the CS. Indeed, the temperature shift due to such a void is only $\Delta T \sim -20 \, \mu K$, whereas the CS shows a temperature decrement of $\Delta T \sim -150 \, \mu K$ at the centre. This was the main argument made by~\cite{Mackenzie:2017ioh}, who concluded, based on the 2dF-VST ATLAS galaxy redshift survey, that the CS may have a primordial origin rather than being imprinted by a void. In standard $\Lambda$CDM a temperature profile similar to the one of the CS is not expected even considering multiple aligned voids~\cite{Naidoo:2015gab}.

The presence of a void can also be detected through gravitational lensing. Indeed, the DES found a lensing signature around the Eridanus region with high confidence through the number count of red galaxies, using the expected galaxy bias to convert it to the dark matter density contrast and to estimate the convergence and gravitational potential. The present data on the lensing and number count of galaxies in that region however does not cover large redshifts, since available galaxy catalogs are still mostly limited to $z\leq1$, and even lower redshifts if high completeness is required. One may thus naturally wonder if  a large enough void could still exist at higher redshifts.

In this work we close this gap, using only the CMB data, by searching for the lensing footprint of a void located anywhere between us and the surface of last scattering (SLS). In~\cite{Masina:2009wt}, the lensing deflection that occurs to a CMB photon when traveling through such structure was shown to affect the CMB power spectrum and bispectrum also at high $\ell$. Lensing introduces correlations in the non-diagonal two-point function, and in~\cite{Masina:2010dc} these were shown to carry a high signal-to-noise ratio, of at least $\mathcal{O}$(10), such that data from the Planck satellite itself should have the capability to confirm or rule out the void explanation for the CS. Since a void would leave both lensing and temperature imprints, one can verify whether they are both compatible with the void conjecture. Here we follow this approach making use of the Planck 2018 data to search for the integrated lensing signal on the CMB from a void that could account for the temperature profile fit (due to ISW and RS effects), using the non-diagonal correlations in harmonic space.

\begin{figure}[t]
    \centering
    \includegraphics[width=0.99\textwidth]{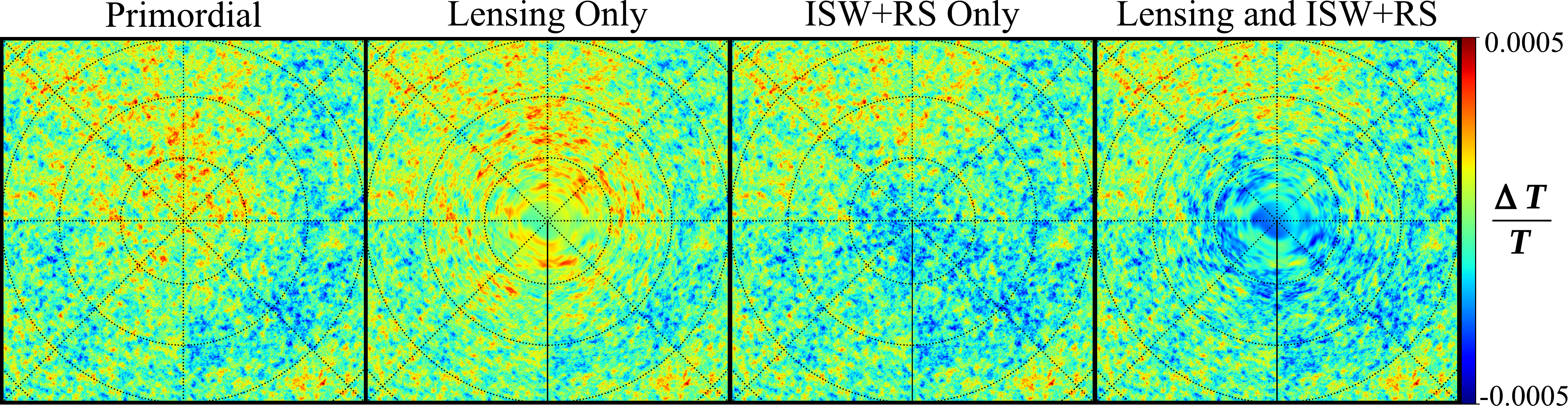}
    \caption{Lensing, ISW and RS effects in temperature maps, in the presence of a void, obtained on a realization of a primordial Gaussian map (first panel) using our HEALPix Void code (see section \ref{sec:Methodology}). All plots span a $30^\circ \times 30^\circ$ range. The second panel includes lensing, artificially enhanced by a factor $\sim50$ from the temperature best fit for clarity. The third combines ISW+RS, artificially enhanced by a factor of 4. The last panel is the total sum.
    }\label{fig:simulations_footprints}
\end{figure}

In Figure~\ref{fig:simulations_footprints} we display the qualitative behavior of each relevant effect: lensing, ISW and RS, using a code which we will describe below.

Finally, we note that the void interpretation of the Cold Spot is further complicated by another claimed anomaly, relating to the amount of ISW effect produced by voids in general. To wit, voids of size $100$ Mpc/$h$ and larger observed from several surveys seem to exhibit a $4.1\pm2.0$ ($5.2\pm1.6$) [$3.6 \pm 2.1$] times larger temperature decrement than predicted by theory using DES Y3 (BOSS+DES Y3) [eBOSS] data, but depending on the used catalog this factor can go up to $\sim 10$~\cite{DES:2018nlb,Kovacs:2021mnf}. This unexplained excess ISW is sometimes encoded into an extra empirical multiplicative parameter $A_{\rm ISW}$, which is unity for $\Lambda$CDM. However this tension was not verified using other methodologies~\cite{Hang:2021kfx}. %\psf{Changing the dark matter equation-of-state it may be possible to explain the $A_{\rm ISW}$ \cite{Naidoo:2022rda}.}
If the ISW effect is enhanced compared to $\Lambda$CDM expectations by a factor $A_{\rm ISW} \sim 5$, the detected void in the CS region could instead account for the CMB temperature anomaly~\cite{Kovacs:2021wnc}.
%,Kovacs:2017hxj,Cai:2016rdk

%%%%%%%%%%%%%-LENSING-%%%%%%%%%%%%%%%%%%
%%%%%%%%%%%%%%%%%%%%%%%%%%%%%%%%%%%%%%%%
%%%%%%%%%%%%%%%%%%%%%%%%%%%%%%%%%%%%%%%%
\section{Lensing effects due to a void} \label{sec:lensing}
%%%%%%%%%%%%%%%%%%%%%%%%%%%%%%%%%%%%%%%%
%%%%%%%%%%%%%%%%%%%%%%%%%%%%%%%%%%%%%%%%

A photon that travels through a void region is subject to deflections, which may be treated accurately using the weak lensing formalism. Following \cite{Masina:2010dc}, we consider a configuration with an observer that looks at the CMB through a spherical void
in the $\hat{z}$ axis direction. We restrict only to the case where the void is completely within the distance to the surface of last scattering and such that the observer is not inside it.

As in \cite{Masina:2008zv}, we describe the effect of a spherical void using the Lemaître-Tolman-Bondi (LTB) metric,
\begin{eqnarray}\label{LTBmetric}
    \dd s^2=-\dd t^2+\frac{R^{\prime 2}(r,t)}{1+2E(r)} \dd r^2+R^2(r,t)(\dd\theta^2+\sin^2\theta \dd\varphi^2) \, ,
\end{eqnarray}
where $^\prime \equiv \dd/\dd r$, $E(r)=-k(r)r^2/2$, $k(r)$ is a curvature function that defines the shape of the void and $R(r,t)=a(t)r$. The LTB metric can be matched to an external flat Friedmann-Lemaître-Robertson-Walker (FLRW) metric, either exactly or asymptotically.  In the first case, the matching is done at a radius $r_L$, by simply setting $k(r>r_L)=0$ and $k'(r=r_L)=0$. In the second case one has to choose a curvature function that goes rapidly to zero after a typical radius, which we may also name $r_L$.  In this paper we will focus on the second possibility, as it gives a better fit of the Planck temperature data through the ISW and RS effects. The void density contrast is defined as $\delta(r)=(\rho(r)-\bar{\rho})/\bar{\rho}$, where $\rho(r)$ is the matter energy density and $\bar{\rho}$ is the average density of the external FLRW universe.

The void can then be characterized by the following parameters: its comoving radius, the comoving distance $D_V$ from us to its centre and  the amplitude of its density contrast at a reference point, which we choose to be the centre ($\delta_0\equiv \delta (r=0)$) at present time (as opposed to the corresponding time inside our past light-cone). In addition, the precise shape of the void density profile has to be chosen.

The observer detects the primordial perturbations on the SLS plus the secondary effects due to this anomalous structure. We write the observed temperature fluctuation as the sum of four components:
\begin{eqnarray}\label{Tfluctuationcom}
    \frac{\Delta T}{T_0} = \frac{\Delta T}{T_0}^{(\rm P)}+\frac{\Delta T}{T_0}^{(\rm RS)}+\frac{\Delta T}{T_0}^{(\rm ISW)}+\frac{\Delta T^{(\rm L)}}{T_0}\,,
\end{eqnarray}
Where $\Delta T^{(i)} \equiv T^{(i)} - T_0$ with  $i={\rm P}, \,{\rm RS}, \,{\rm ISW}, \,{\rm L}$ and $T_0$ it the total observed monopole.

%%%%%%%%%%%%%%%%%%%%%%%%%%%%%%%%%%%%%%%%
%%%%%%%%%%%%%%%%%%%%%%%%%%%%%%%%%%%%%%%%
\subsection{Lensing Profile}\label{subsec: L profile}
%%%%%%%%%%%%%%%%%%%%%%%%%%%%%%%%%%%%%%%%
%%%%%%%%%%%%%%%%%%%%%%%%%%%%%%%%%%%%%%%%

We may compute the perturbed paths of light rays from the CMB to the observer by treating the LTB inhomogeneity as a perturbation of a FLRW metric with  a gravitational potential~$\Phi$. The metric in~\eqref{LTBmetric} may be perturbatively written as~\citep[see, e.g.,][]{Masina:2008zv, Biswas:2007gi}
\begin{eqnarray}\label{Newtonianline}
    \dd s^2  =  a^2(\tau)\big[-(1+2\Phi)d\tau^2+(1-2\Phi)\dd x^i \dd x^j\big],
\end{eqnarray}
where $\tau$ is the conformal FLRW time, $a$ the scale factor normalized to 1 at present, and $x^i$ is a dimensionless comoving coordinate. In~\cite{Biswas:2007gi} only matter domination was considered, which leads to a gravitational potential constant in time and related to the curvature function by:
\begin{eqnarray}\label{Phi_r}
    \Phi_e(r)=\frac{3}{10}\int^{r_L}_{r}k(\bar{r})\bar{r}d \bar{r} \; .
\end{eqnarray}
Such a form is still approximately true in $\Lambda$CDM at sufficiently early times, when the cosmological constant is negligible.
We employ a simple Gaussian function to model the void, namely
\begin{eqnarray}\label{k_r}
    k(r)=k_0 \exp \Big[-\frac{1}{2}\Big(\frac{r}{r_V}\Big)^2\Big] \, .
\end{eqnarray}
We chose this form, which was also used in~\cite{Nadathur:2014tfa, Finelli:2014yha}, because it gives a good fit to the shape of the temperature profile of the CS.

The void in practice extends to a size of a few standard deviations $r_V$, after which the gravitational potential $\Phi$ becomes negligible.  For definiteness, we set the integration limit as $r_{L}\equiv 4.2 r_V$ throughout. This corresponds to a region large enough to encompass 99\% of the void mass distribution, as described in Figure~\ref{fig:P_theta} in Appendix~\ref{appendix:sims}. We find that any value around this value or greater leads to negligible changes.
The angle subtended by $r_V$ is defined as $\theta_V$, with $\tan\theta_V=r_V/D_V$. I.e., $\theta_V$ is the angular radius corresponding to one standard deviation in the curvature profile. For reference the best-fit angular size of the Cold Spot considering the Gaussian profile is $\theta_V \sim 6.5^\circ$, while and the DES Eridanus supervoid is roughly adjusted with $\theta_V \sim 9^\circ$~\cite{Kovacs:2021wnc}.

With the addition of a cosmological constant,  $\Phi$ becomes time-dependent and, using the linear growth approximation, is given by
\begin{eqnarray}\label{eq:lin-growth}
    \Phi(r,\tau)=\frac{D_+(a)}{a}\Phi_{e}(r), \;\;\;\;\;\;  D_+(a)=\frac{5\Omega_{m0}}{2}\frac{H(a)}{H_0}\int_0^a\frac{da'}{(a'H(a')/H_0)^3}.
\end{eqnarray}
where $H(a)=H_0\sqrt{\Omega_{m0}a^{-3}+(1-\Omega_{m0})}$ is the Hubble parameter, $H_0$ the Hubble constant, $\Omega_{m0}$ the present matter density and $D_+(a)$ is the linear growth factor. It is defined such that $\delta(a)=\delta(1) D_+(a)/D_+(1)$ and normalized to $D_+(a) = a$ during matter domination. With the inclusion of a cosmological constant, $k_0$ can be obtained by inverting the Poisson equation $\nabla^2\Phi_e(r)=(3/2)\Omega_{m0}H_0^2\delta a/D_+(a)$, giving:
\begin{eqnarray}\label{k0}
     k_0 =\frac{5}{3}\frac{\Omega_{m0} H_0^2\delta_0}{D_+(1)}\,.
\end{eqnarray}

Once the gravitational potential is defined, the lensing effect on the temperature along the line of sight, $\Delta T^{(\rm L)}/T$, can be computed in a gradient expansion as a function of the unlensed fluctuation \cite{Masina:2009wt} in a given direction $\hat{n}$. This lensing contribution is given by
\begin{eqnarray}\label{Lfluctuation}
    \frac{\Delta T^{(\rm L)}}{T_0}(\hat{n}) = \partial _i\frac{\Delta T^{(\rm P)}}{T_0}(\hat{n})\partial ^i\Theta(\hat{n})+\partial _j\partial _i\frac{\Delta T^{(\rm P)}}{T_0}(\hat{n})\partial ^j\Theta(\hat{n})\partial ^i\Theta(\hat{n})+\cdots \,,
\end{eqnarray}
where we will consider only up to the first order perturbative term and $\Theta$ is the so-called lensing potential. The observed deflection vector is given by the standard weak lensing equation, as an integral along the photon path,
\begin{eqnarray}\label{lensing_potential}
    \nabla_{\hat{n}}\Theta=-2\int^{\tau_{0}}_{\tau_{\rm SLS}} d\tau \frac{\tau-\tau_{\rm SLS}}{(\tau_0-\tau_{\rm SLS})(\tau_0-\tau)}\nabla_{\hat{n}}\Phi,
\end{eqnarray}
where  $\tau_0$ and $\tau_{\rm SLS}$ are the conformal time at the observer and at the SLS respectively, $\nabla_{\hat{n}}$ is the angular derivative and $\tau$ has an analytical expression given as:
\begin{eqnarray}\label{tauofa}
    \tau(a)=\int_0^a\frac{d\Tilde{a}}{\Tilde{a}^2H(\Tilde{a})}=\frac{2\sqrt{a}}{H_0\sqrt{\Omega_{m0}}} \, {}_2F_1\bigg[\frac{1}{2},\frac{1}{6},\frac{7}{6},\frac{\Omega_{m0}-1}{\Omega_{m0}}a^3\bigg].
\end{eqnarray}
Given a temperature anisotropy $\Delta T^{(i)}(\hat{n})/T$ and a lensing profile $\Theta(\hat{n})$, their spherical decompositions are respectively:
\begin{eqnarray}\label{alm_blm}
    a^{(i)}_{\ell m} \equiv \int d \hat{n} \frac{\Delta T^{(i)}(\hat{n})}{T_0} Y^*_{\ell m}(\hat{n}) \ ,  \hspace*{5mm}     \ b_{\ell m} \equiv \int d \hat{n} \Theta(\hat{n})  Y^*_{\ell m}(\hat{n})\,.
\end{eqnarray}
This leads to the standard equation for the lensed $a_{\ell m}$:
\begin{eqnarray}\label{alm_w_lens}
    a_{\ell m}=a_{\ell m}^{(\rm P)}+a_{\ell m}^{(\rm L)}\,,
\end{eqnarray}
where $a_{\ell m}^{(\rm P)}$ is the unlensed (primordial) signal and $a_{\ell m}^{(\rm L)}$ is the lensing correction, given by
\begin{eqnarray}\label{alm_w_lens_def}
    a_{\ell_1m}^{(\rm L)}=\sum_{\ell_2,\ell_3} (-1)^m \, G_{\hspace{2mm}\ell_1 \hspace{2mm} \ell_2 \hspace{1mm} \ell_3}^{-m \ m \  0}  \ \frac{\ell_2(\ell_2+1)-\ell_1(\ell_1+1)+\ell_3(\ell_3+1)}{2} a_{\ell_2m}^{(\rm P)}b_{\ell_30},
\end{eqnarray}
where we introduced the Gaunt integrals, given in terms of the Wigner 3-j symbols, as follows:
\begin{equation}\label{Gaunt_integrals}
    G_{\ell_1 \hspace{2mm}\ell_2 \hspace{2mm} \ell_3}^{m_1 m_2 m_3} \equiv \sqrt{\frac{(2\ell_1+1)(2\ell_2+1)(2\ell_3+1)}{4 \pi}} \left({\begin{array}{ccc} \ell_1 & \ell_2 & \ell_3 \\ 0 & 0& 0 \end{array}}\right) \left({\begin{array}{ccc} \ell_1 & \ell_2 & \ell_3 \\ m_1 & m_2& m_3 \end{array}}\right)\,.
\end{equation}

\begin{figure}[t]
    \centering
    \includegraphics[width=\textwidth]{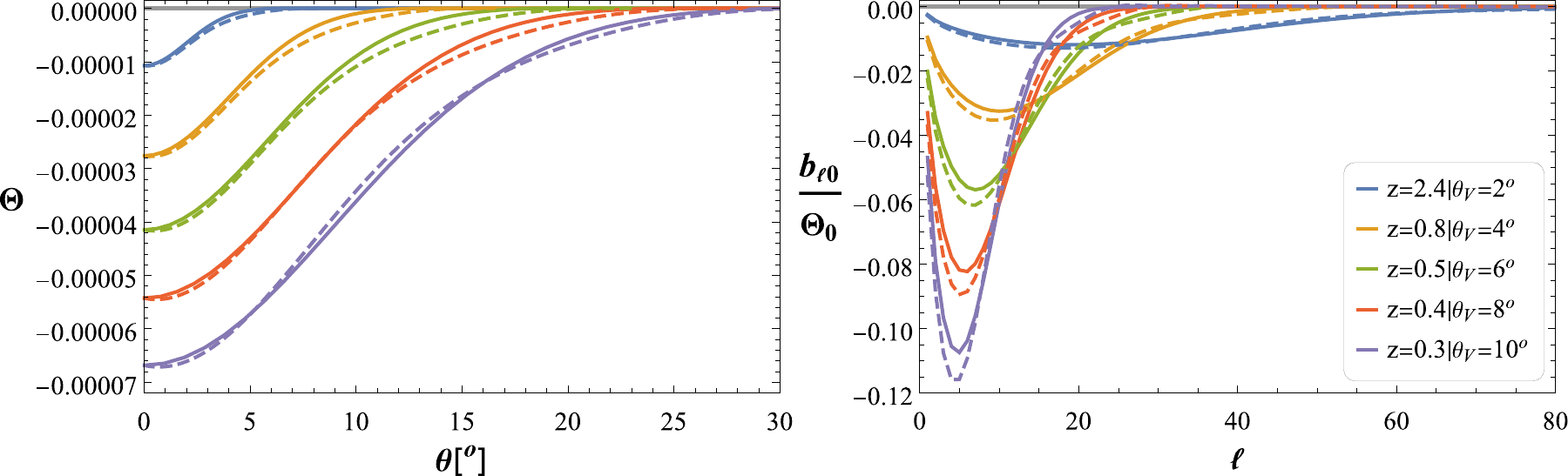}
    \caption{\emph{Left:} lensing potential profile generated by a fiducial void with $\delta_0=-0.25$, $r_V=140$Mpc/$h$. \emph{Right:} corresponding $b_{\ell0}/\Theta_0$ as a function of the multipole $\ell$. We show both plots for various redshifts and angular sizes. The solid (dashed) lines are calculated using our (observed DES) $k(r)$ profile in Eq.~\eqref{k_r}. }\label{fig:theta_vs_bl_combined}
\end{figure}

Solving equation \eqref{lensing_potential} and using spherical coordinates, $0\leq \theta \leq \pi$ and $0\leq \phi < 2\pi$, we parameterized the lensing potential as
\begin{eqnarray}\label{parameterizeLensing}
    \Theta(\theta,\phi)= \Theta_0p(\theta) \quad {\rm for} \quad \theta < \theta_L\,,
\end{eqnarray}
where  $\Theta_0$ is the amplitude and $p(\theta)$ is the profile function, obtained through the fitting of a numerical solution of equation \eqref{lensing_potential} (see appendix \ref{appendix:sims}), and normalized so that $p(0)\equiv-1$. The angular integration limit $\theta_L$ is conservatively set as $\arctan(r_L/D_V)$. We also set $p(\theta\geq\theta_L)=0$. The profile has no dependence on $\phi$ because we choose the $\hat{z}$ axis to point towards the center of the void; as a consequence the only non-vanishing $b_{\ell m}$ are the $b_{\ell 0}$, which are real.
Since the $b_{\ell 0}$ are linear in $\Theta_0$ it is convenient to define
\begin{eqnarray}\label{bl0_1}
    b_{\ell 0} \equiv  \Theta_0 \hat{b}_{\ell 0} ,
\end{eqnarray}
where $\hat{b}_{\ell 0} \equiv 2\pi\int d\theta \sin\theta   \, p(\theta) \, Y_{\ell 0}(\theta)$. In Figure~\ref{fig:theta_vs_bl_combined} we depict the lensing potential and its respective spherical harmonics components expansion for some cases.

In order to test whether our results are very sensitive to the chosen curvature profile, we also computed results using a data driven profile shape. To wit, we employed the typical lensing potential from a catalog containing $\sim3600$ voids observed by DES~\citep{DES:2022uvb} which provides the average convergence as a function of $\theta$. We computed $\Theta(\hat n)$ from the average lensing convergence $\kappa$, solving the equation
\begin{eqnarray}\label{eq:convergence}
    \kappa(\hat n) = \frac{1}{2}\nabla^2\Theta(\hat n) \,,
\end{eqnarray}
assuming that $\Theta(\hat n)$ goes to zero far outside the void. This new lensing potential turns out to be very similar to our choice Eq.~\eqref{k_r}, see Figure~\ref{fig:theta_vs_bl_combined}. As will be shown below, simulating data with either profile resulted in very similar constraints.

%%%%%%%%%%%%%%%%%%%%%%%%%%%%%%%%%%%%%%%%
%%%%%%%%%%%%%%%%%%%%%%%%%%%%%%%%%%%%%%%%
\subsection{Non-diagonal two-point function}\label{subsec: Two Point Function}
%%%%%%%%%%%%%%%%%%%%%%%%%%%%%%%%%%%%%%%%
%%%%%%%%%%%%%%%%%%%%%%%%%%%%%%%%%%%%%%%%
As already shown in \cite{Masina:2010dc}, a void that accounts for the CS leads to a significant signal in the non-diagonal part of the two point correlation function of the temperature fluctuations. From Eq.~\eqref{alm_w_lens_def}, we can write the first order contribution to the two-point correlation function due  to the primordial and lensing temperature fluctuations as follows:

\begin{equation}\label{2point_PL}
    \big\langle a_{\ell_1 m_1}^{(\rm P)} a_{\ell_2 m_2}^{(\rm L)*}  \big\rangle =  \Theta_0\delta_{m_1m_2}(-1)^{m_2} C_{\ell_1}^{(\rm P)}
    \sum_{\ell_3}G_{\ell_1 \, \ell_2 \, \ell_3}^{-m_2 \, m_2 \,  0} \, \frac{\ell_2(\ell_2+1)-\ell_1(\ell_1+1)+\ell_3(\ell_3+1)}{2} \hat{b}_{\ell_30}\, ,
\end{equation}
using the fact that for a primordial isotropic and Gaussian signal the two-point correlation functions are given by $\big\langle a_{\ell_1 m_1}^{(\rm P)} a_{\ell_2 m_2}^{(\rm P)*}\big\rangle = \delta_{\ell_1 \ell_2}\delta_{m_1 m_2} C_{\ell_1}^{(\rm P)}$. Clearly the contribution is diagonal in the index $m$. The diagonal part in the $\ell$ index, $\ell_1=\ell_2$, vanishes due to properties of the Gaunt integrals~\cite{Masina:2009wt}. On the other hand, the off-diagonal part leads to a statistically significant signal, because although individually small, the correlations between $\ell_1$ and $\ell_2$ extend far outside the diagonal.
From Figure \ref{fig:theta_vs_bl_combined} it is clear that most of the signal is contained in $|\ell_1-\ell_2|\equiv \Delta\ell\leq 40$, so we will use this range for our lensing estimator as a good compromise between computational time and accuracy (see section \ref{sec:lens estimator} and Appendix \ref{appendix:bias_plot}).

Quantitatively, we can construct an estimator by defining the quantity
\begin{eqnarray}\label{Flm}
    f_{\ell_1\ell_2m}\equiv \frac{1}{2}(a_{\ell_1m}^*a_{\ell_2m}+a_{\ell_1m}a_{\ell_2m}^*)\,.
\end{eqnarray}
Taking its statistical average we get, to first order:
\begin{eqnarray}\label{flm}
    f_{\ell_1 \ell_2 m}^{\rm TH} \equiv \big\langle f_{\ell_1\ell_2m}\big\rangle  =  \big\langle a_{\ell_1m}^{(\rm P)*}a_{\ell_2m}^{(\rm L)}\big\rangle +\big\langle a_{\ell_1m}^{(\rm L)*}a_{\ell_2m}^{(\rm P)}\big\rangle \,.
\end{eqnarray}
Using the Gaunt integral property $G_{\ell_1 \, \ell_2 \, \ell_3}^{-m \, m \,  0}= G_{\ell_2  \,\ell_1 \, \ell_3}^{-m \, m \,  0}$, we have
\begin{equation}\label{fTH}
\begin{split}
    f_{\ell_1 \ell_2 m}^{\rm TH} = \Theta_0(-1)^m \sum_{\ell_3}G_{\ell_1 \, \ell_2 \, \ell_3}^{-m \, m \,  0}& \bigg( C_{\ell_1}^{(\rm P)} \ \frac{\ell_1(\ell_1+1)-\ell_2(\ell_2+1)+\ell_3(\ell_3+1)}{2} \\ & + C_{\ell_2}^{(\rm P)} \ \frac{\ell_2(\ell_2+1)-\ell_1(\ell_1+1)+\ell_3(\ell_3+1)}{2} \bigg) \hat{b}_{\ell_30}\,.
\end{split}
\end{equation}
Since the variance of $ f_{\ell_1\ell_2m}$  is $\sigma _{f_{\ell_1 \ell_2 m}}^2=\frac{1}{2}C_{\ell_1}^{(\rm P)}C_{\ell_2}^{(\rm P)}(1+\delta _{m0})$, we can define a S/N ratio as:
\begin{eqnarray}\label{StoN}
    \left(\frac{S}{N}\right)^2= (2-\delta_{m0}) \sum_{\ell_1 = 2}^{\ell_{\rm max}}  \sum_{\ell_2=\ell_1+1}^{\ell_1+\Delta \ell} \sum_{m=0}^{\ell_1} \frac{\big[f_{\ell_1 \ell_2 m}^{\rm TH}\big]^2  }{\sigma _{f_{\ell_1 \ell_2 m}}^2}= (2-\delta_{m0})\sum _{\ell_1,\ell_2,m} \frac{\big[f_{\ell_1 \ell_2 m}^{\rm TH}\big]^2}{C_{\ell_1}^{(\rm P)}C_{\ell_2}^{(\rm P)}}\,,
\end{eqnarray}
which is a function of $\ell_{\rm max}$.

%%%%%%%%%%%%%%%%%%%%%%%%%%%%%%%%%%%%%%%%
%%%%%%%%%%%%%%%%%%%%%%%%%%%%%%%%%%%%%%%%
\section{Temperature anisotropy due to a void}\label{subsec: T profile}
%%%%%%%%%%%%%%%%%%%%%%%%%%%%%%%%%%%%%%%%
%%%%%%%%%%%%%%%%%%%%%%%%%%%%%%%%%%%%%%%%
\subsection{Linear and non linear ISW}

The total temperature shift for a photon passing through the void is calculated here using perturbation theory in $\Lambda$CDM~\cite{Tomita:2007db,Nadathur:2014tfa}. We consider first and second order perturbations around FLRW, with the line element
\begin{eqnarray}\label{line_element}
    \dd s^2 = g_{\mu\nu}\dd x^\mu \dd x^\nu= a^2(\tau)(g_{00}\dd x^0 \dd x^0+g_{0i} \dd x^0 \dd x^i+g_{ij} \dd x^i \dd x^j),
\end{eqnarray}
where the metric can be written as
\begin{align}\label{gij}
    g_{00} &\;=-\Big(1+2\psi^{(1)}+\psi^{(2)} \Big), \\
    g_{0i} &\;=z_i^{(1)}+\frac{1}{2}z_i^{(2)}, \\
    g_{ij} &\;=\big(1-2\phi^{(1)}-\phi^{(2)}\big)\delta_{ij}+\chi_{ij}^{(1)}+\frac{1}{2}\chi_{ij}^{(2)},
\end{align}
and $\psi^{(n)}$, $\phi^{(n)}$, $z_i^{(n)}$, and $\chi_{ij}^{(n)}$ are the $n$-th order metric perturbations. The first order temperature fluctuation of photons emitted from the position $\varepsilon$ and observed at $o$ is
\begin{eqnarray}\label{deltaT_1st_order}
    \frac{\Delta T^{(1)}}{T_0} = \psi_\varepsilon^{(1)}-\psi_o^{(1)}+ \Big(v_o^{(1)i}-v_\varepsilon^{(1)i}\Big) e_i+\eta+I_{1}\,,
\end{eqnarray}
with $v^{(1)i}$ being the velocity perturbation and $\eta$ the temperature anisotropy at the emission. The term $I_{1}$ is an integrated term that represent the ISW contribution
\begin{eqnarray}\label{deltaT_ISW}
  \frac{\Delta T}{T_0}^{\rm ISW}=I_{1}=-\int_{\tau_\varepsilon}^{\tau_o}\dd\tau \Big(\dot{\psi}^{(1)}+\dot{\phi}^{(1)}\Big),
\end{eqnarray}
where a dot denotes the derivative with respect to conformal time $\tau$. In our case only scalar linear perturbations are present. Similarly, the corresponding expression for the second-order anisotropy, RS term, is
\begin{eqnarray}\label{deltaT_RS}
    \frac{\Delta T}{T_0}^{\rm RS}=\frac{1}{2} \int_{\tau_\varepsilon}^{\tau_o}\dd\tau \Big(\dot{\psi}^{(2)}+\dot{\phi}^{(2)}+[I_{1}]^2\Big),
\end{eqnarray}
where we dropped second-order vector and tensor terms, which are suppressed for sub-horizon sized structures~\cite{Matarrese:1997ay}. The ISW squared term is also negligible and can be dropped.

Solving the Einstein equations one can obtain the solutions for the first and second order perturbations in the growing mode, in terms of $\Phi_{e}(r)$ in~\eqref{Phi_r}, as~\cite{Tomita:2005et,Tomita:2007db}
 \begin{align}
 \psi^{(1)} &\;=\phi^{(1)}=\frac{5}{3}\bigg(1-\frac{\dot{a}}{a}\dot{P}\bigg)\Phi_{e} ,  \label{phi1} \\
    \psi^{(2)} &\;=\phi^{(2)}= \frac{100}{9}(\zeta_1\Phi_{e,i}\Phi_{e,i}+\frac{9}{2}\zeta_2\Upsilon_0)\,.
    \label{phi2}
\end{align}
where we defined:
\begin{eqnarray}\label{P}
    P(\tau)=\int_0^\tau \dd\Tilde{\tau} a^{-2}(\Tilde{\tau})\int_0^{\Tilde{\tau}}\dd\bar{\tau} a^{2}(\bar{\tau}),
\end{eqnarray}
which is determined by the background cosmology. Note that the growth functions $D_+$ and $P$ are related by $D_+(a)/a = (5/3)\big[1-({\dot{a}/a})\dot{P}\big]$. Additionally, $\Phi_{e,i}$ stands for $\partial \Phi_e/\partial r^i$, and $\Upsilon_0$ is a non-local quantity \cite{Biswas:2007gi,Nadathur:2014tfa}, which is given in terms of $\Phi_e$ as
\begin{eqnarray}\label{Upsilon_0}
    \nabla^2\Upsilon_0 = \Phi_{e,ij}\Phi_{e,ij} - \big( \nabla^2 \Phi_e \big)^2 \,,
\end{eqnarray}
and~\cite{Tomita:2007db}
\begin{eqnarray}\label{zeta_1}
    \zeta_1 &=&\frac{1}{4}P\bigg(1-\frac{\dot{a}}{a}\dot{P}\bigg) \,,
    \nonumber \\
    \zeta_2 &=&\frac{1}{21}\frac{\dot{a}}{a}\bigg(P\dot{P}-\frac{\dot{Q}}{6}\bigg)-\frac{1}{18}\bigg(P+\frac{(\dot{P})^2}{2}\bigg)\,,
    \nonumber \\
    Q &=& \int_0^{\tau}\dd\Tilde{\tau} a^{-2}(\Tilde{\tau})\int_0^{\Tilde{\tau}}\dd\bar{\tau} a^2(\bar{\tau})\bigg[-P(\bar{\tau})+\frac{5}{2}\big(\dot{P}(\bar{\tau})\big)^2\bigg].
\end{eqnarray}

With all the necessary functions defined one may use equations \eqref{deltaT_ISW}, \eqref{phi1} and \eqref{P} to obtain the ISW contribution as a function of the angle $\theta$ from the center of the void as
\begin{eqnarray}\label{ISWintheta}
  \frac{\Delta T^{\rm ISW}}{T_0}(\theta) =  \int_0^{z_{\rm SLS}}\bigg[\bigg(\frac{\ddot{a}}{a}-3\frac{\dot{a}^{2}}{a^2}\bigg)\dot{P}+\frac{\dot{a}}{a}\bigg]\frac{k_0 r_V^2 \,\exp{\Big[-\frac{1}{2}\frac{\Tilde{r}^2(z,\theta)}{r_V^2}\Big]}}{H(z)}\dd z,
\end{eqnarray}
where $z_{\rm SLS} = 1090$ and where we computed the integral on a straight line defined as $\Tilde{r}(z,\theta)=\big[\big(r(z_V)-r(z)\cos{(\theta)}\big)^2+r(z)^2\sin{(\theta)}^2]^{1/2}$, with $r(z)=\tau_0-\tau(z)$ being the comoving distance to redshift $z$. Similarly, using equations \eqref{deltaT_RS}, \eqref{phi2} and \eqref{Upsilon_0} the RS contribution becomes
\begin{eqnarray}\label{RSintheta}
   \frac{\Delta T^{\rm RS}}{T_0}(\theta) = \frac{1}{2} \int_0^{z_{\rm SLS}}\bigg(2\frac{\Tilde{r}^2(z,\theta)}{r_V^2}\dot{\zeta}_1+9\dot{\zeta}_2\bigg)\frac{(k_0r_V)^2 \,\exp{\Big[-\frac{\Tilde{r}^2(z,\theta)}{r_V^2}\Big]}}{H(z)} \dd z \,.
\end{eqnarray}

\begin{figure}[t]
    \centering
    \includegraphics[width=0.65\textwidth]{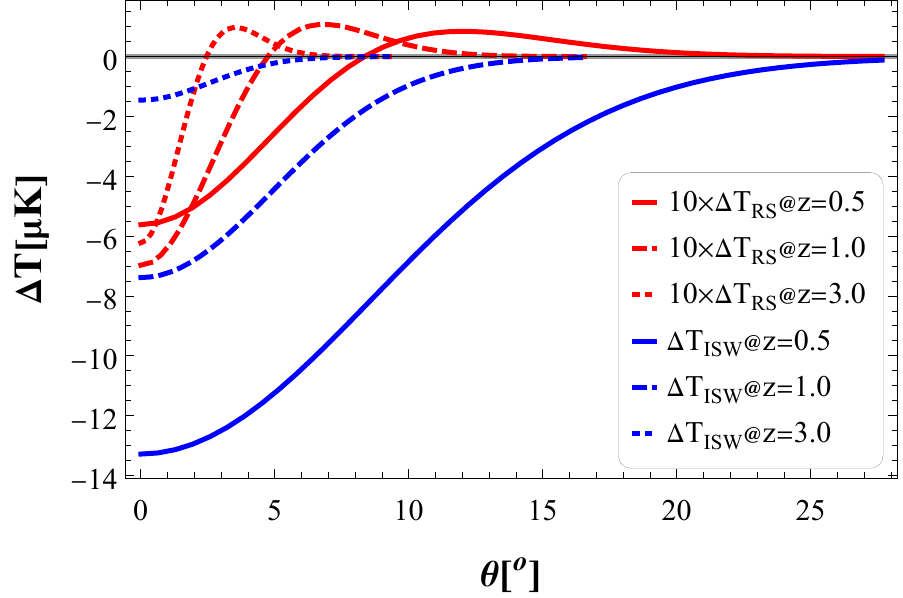}
    \caption{The profile of ISW (blue curves) and RS (red curves) temperature profiles, for an LTB void located at various redshifts, with $r_V=140$Mpc/$h$ and $\delta_0 =-0.25$. The RS effect is amplified by a factor of 10 for clarity.}\label{fig:deltaT_RS_ISW}
\end{figure}

In Figure \ref{fig:deltaT_RS_ISW} we show an example of the angular dependence of temperature anisotropy profiles $\Delta T(\theta)$, which demonstrates that the RS introduces a hot ring around the CS and is more relevant for high $z$.  In Figure~\ref{deltaT_amplitudes} in Appendix~\ref{app:RS} we also plot the ratio between ISW and RS contributions for a wider range of $z$ and $\delta_0$ values.

%\pagebreak
%%%%%%%%%%%%%%%%%%%%%%%%%%%%%%%%%%%%%%%%%%%%%%%
\section{Methodology}\label{sec:Methodology}
%%%%%%%%%%%%%%%%%%%%%%%%%%%%%%%%%%%%%%%%

%%%%%%%%%%%%%%%%%%%%%%%%%%%%%%%%%%%%%%%%
%%%%%%%%%%%%%%%%%%%%%%%%%%%%%%%%%%%%%%%%
\subsection{Estimators} \label{sec:lens estimator}
%%%%%%%%%%%%%%%%%%%%%%%%%%%%%%%%%%%%%%%%
%%%%%%%%%%%%%%%%%%%%%%%%%%%%%%%%%%%%%%%%

Here we present our $\chi^2$ for the lensing and temperature imprints due to a void. For lensing it is a function of two parameters, $\Theta_0$ and $\theta_V$:
\begin{eqnarray}\label{chisq_lensing}
   \chi^2(\Theta_0,\theta_V) =  (2-\delta_{m0})\sum_{\ell_1 = 2}^{\ell_{\rm max}}  \sum_{\ell_2=\ell_1 + 1}^{\ell_1 + \Delta \ell} \sum_{m=0}^{\ell_1} \frac{\left[f^{\rm OBS}_{\ell_1 \ell_2 m} - f^{\rm TH}_{\ell_1 \ell_2 m}(\Theta_0,\,\theta_V)\right]^2}{ C_{\ell_1}^{(\rm P)} C_{\ell_2}^{(\rm P)}} \;,
\end{eqnarray}
where $f_{\ell_1\ell_2 m}^{\rm OBS}$ is given by~\eqref{Flm} for the observed data and $f_{\ell_1\ell_2 m}^{\rm TH}$ by  \eqref{fTH}. We will also consider the effects of noise and masking, which imply the change $C_\ell \rightarrow \Tilde{C}_\ell + \Tilde{N_\ell}$ on equations \eqref{fTH} and \eqref{chisq_lensing}, where $N_\ell$ is the noise power spectrum and a tilde stands for the masked spectra. However, this naïve isotropic $\chi^2$ ignores the  anisotropy introduced by the mask and the noise, which leads to biases that need to be removed in the analysis. The parameters $\Theta_0$ and $\theta_V$ are finally translated to physical parameters  $\delta_0$, $r_V$ and $D_V$. More details are given in Appendix~\ref{appendix:bias_plot}.

For the temperature fit we computed the following $\chi^2$ as a function of $\delta_0, r_V, D_V$:
\begin{eqnarray}\label{chi_temperature}
    \chi^2(\delta_0, r_V, D_V)=\sum_i\bigg[\frac{\Delta T^{\rm TH}(\theta_i,\delta_0, r_V, D_V)-\Delta T^{\rm OBS}(\theta_i)}{\sigma_{\Delta T^{\rm OBS}(\theta_i)}}\bigg]^2,
\end{eqnarray}
where $\Delta T^{\rm TH}=\Delta T^{\rm ISW}+\Delta T^{\rm RS}$ are the theoretical predictions given by equations \eqref{ISWintheta} and \eqref{RSintheta}, with $\theta_i$ being polar angle of the rings of width $2^\circ$ weighted by the number of pixels. The rings start from the center of the CS up to an angle of $24^\circ$, $\Delta T^{\rm OBS}(\theta_i)$ is averaged over the pixels contained in ring $i$, and $\sigma_{\Delta T^{\rm OBS}(\theta_i)}$ is the standard deviation of $\Delta T^{\rm OBS}(\theta_i)$ over many realistic end-to-end (E2E) simulations provided by the Planck Collaboration for CMB and noise~\citep{Aghanim:2018fcm}.

\subsection{Pipeline summary}

In this section we summarize the main ideas, while more details of the pipeline, bias removal and lensing potential fitting are discussed in appendix~\ref{appendix:pipeline}. We have two paths in our pipeline, one for temperature and another one for lensing.

The temperature path is the simplest one. We started with both Planck 2018 maps. We rotated these maps in harmonic space, using the HEALPix package~\cite{Gorski:2004by}, so that the CS direction points towards the north pole and computed the average $\Delta T$ for 12 rings of pixels with width $2^\circ$ centered in the CS. For our Gaussian void profile we computed the $\chi^2$ varying $\delta_0$, $r_V$ and $D_V$ as in equation~\eqref{chi_temperature}. The temperature estimator does not have significant biases due to the mask or to the anisotropic noise, so the outputs are the final results.

For lensing, the estimator will have intrinsic biases due to the anisotropic noise and the mask, so the pipeline is less straightforward. We employed a pipeline which is an extension of the one used in~\cite{Ferreira:2021omv} to measure off-diagonal correlations due to peculiar velocity effects, which introduce $\ell, \ell+1$ correlations~\cite{Notari:2011sb}. In particular, because it removes the Dipole Distortion (see~\cite{Notari:2015daa}) which could only be modelled for SMICA and NILC, it is restricted to these two component separation maps. We analyzed both in order to check for possible systematics.

We first generated mock simulations to train the estimator. We started with the E2E CMB simulations and applied the ISW, RS and lensing effects introduced by a void, using our modified HEALPix code dubbed HEALPix Void.\footnote{\url{https://github.com/pdsferreira/healpix_void}} The code capabilities were illustrated in Figure~\ref{fig:simulations_footprints}. We then summed the theoretical results with the noise E2E simulations, which included a combination of beaming, aberration and Doppler effects due to the solar system velocity including the Dipole Distortion, systematics, foregrounds, exposure time and lensing from other sources (through independent lensing potential realizations)~\cite{Aghanim:2018fcm}. We then applied the mask to the resulting map, rotated it so that the CS direction points to the north pole and expanded it in spherical harmonics.

We repeated this for $\sim 53000$ simulations over a grid of $1550$ different combinations of $\Theta_0$ and $\theta_V$, for both SMICA and NILC maps. We estimated $\Theta_0$ and $\theta_V$ on these simulations and used it to train the bias removal step using two methods, a Random Forest Regression (RFR)~\cite{breiman2001random,Pedregosa:2011ork} and a $3^{\rm rd}$ order polynomial fit on $\Theta_0$ and $\theta_V$. We tested the trained bias-removal using 10000 extra simulations with random values of $\Theta_0$ and $\theta_V$, and 6000 null simulations (without a void) to avoid over-fitting, verifying that both methods work well. We found significantly better bias removal for the RFR method, which we chose as our fiducial tool. The trade-off is that it provides noisier confidence-level contours. Finally we computed the $\chi^2$ for a grid of values for $\Theta_0$ and $\theta_V$ in the CMB observed map, removed the bias and converted from $\Theta_0$ and $\theta_V$ to the physical parameters $r_V$, $D_V$ and $\delta_0$ by using Eq.~\eqref{lensing_potential}. We thus obtained the final results for lensing.

We employed the lensing pipeline also for 112 different directions in a radius of $5.5^\circ$ around the CS canonical direction. These 112 directions are the pixels obtained using HEALPix with NSIDE=64. We found no significantly different results for lensing with respect to the canonical direction.

%%%%%%%%%%%%%%%%%%%%%%%%%%%%%%%%%%%%%%%%
%%%%%%%%%%%%%%%%%%%%%%%%%%%%%%%%%%%%%%%%

%\pagebreak

%%%%%%%%%%%%%%%%%%%%%%%%%%%%%%%%%%%%%%%%%
%%%%%%%%%%%%%%%%%%%%%%%%%%%%%%%%%%%%%%%%%
\section{Main results} \label{sec:results}
%%%%%%%%%%%%%%%%%%%%%%%%%%%%%%%%%%%%%%%%%
%%%%%%%%%%%%%%%%%%%%%%%%%%%%%%%%%%%%%%%%%

We start by discussing the adopted priors in our analysis. We adopted three independent priors. The first one has a practical motivation: to keep computational cost manageable and the bias removal method working well, we limited a maximum value of $\theta_V$, to wit $\theta_V \le 10.1^\circ$. This is over one and a half the best fit size for the temperature likelihood data, and larger than the size needed for the DES Eridanus supervoid. The second prior has to do with how far we allow the void to be. Since we assumed the whole void is in between the SLS and us, we imposed that  $D_V + 3r_V \le D_{\rm SLS}$. I.e., we guaranteed that the center and up to at least the $3\sigma$ of the curvature profile is closer than the center of the SLS. The third prior is $-1\leq \delta_0\leq 0 $, meaning that we only considered voids. In practice, in order to compute this prior we assume $\Omega_{m0}=0.3$ and $H_0 = 70$km/s/Mpc, but the results are very insensitive to this choice of background cosmology.

\begin{figure}[t]
    \centering
    \includegraphics[width=0.65\linewidth]{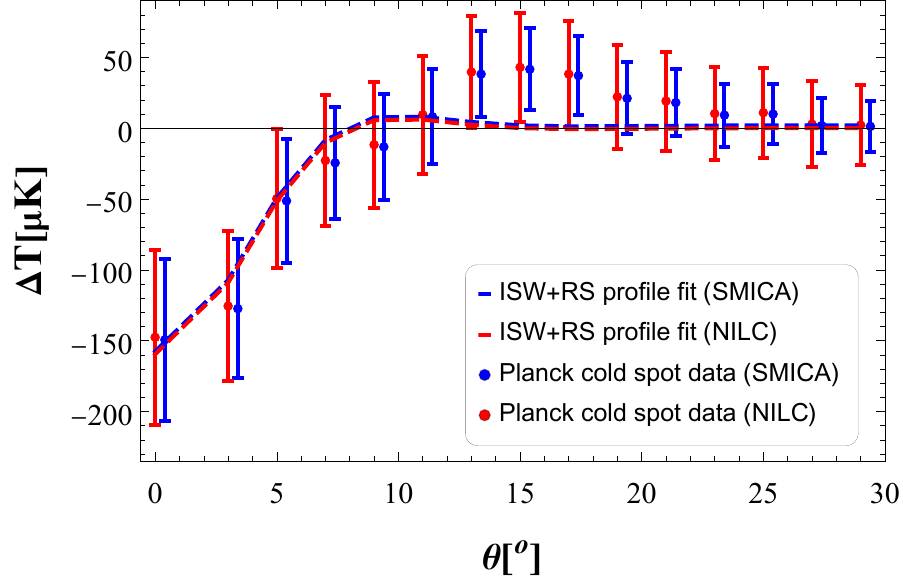}
    \caption{Planck 2018 SMICA and NILC CS temperature data, with best fit theoretical profiles. The data points and error bars are slightly displaced horizontally for clarity.
    }\label{fig:deltaT_obs}
\end{figure}

For the temperature estimator, the best fit values are obtained by minimizing the $\chi^2$ over a grid of values for all the parameters, considering the same grid as the lensing $\chi^2$ after conversion to physical parameters, as per Appendix \ref{appendix:bias_plot}. For both SMICA and NILC we found the same best fit values:  ($\delta_0=-0.34$, $r_V=806\,$Mpc/$h$, $D_V=6910\,$Mpc/$h$). The corresponding theoretical best fit profiles for both maps are shown in Figure~\ref{fig:deltaT_obs}. The expected S/N from the lensing effect of a void described by the above best fit values is $\sim 12$ (see Figure~\ref{fig:StoN}) for $\ell_{{\rm max}}=1600$ and the maximum non-diagonal correlation $\ell_2 - \ell_1 = 40$. For lensing instead we obtained results consistent with no lensing signal. This is in agreement with~\cite{Jakubec:2020eui}, which failed to find any significant anomalous signals in the CMB lensing maps. The best fits parameters were found to be: ($\delta_0=0.00$, $r_V=382\,$Mpc/$h$, $D_V=2520\,$Mpc/$h$) for SMICA and ($\delta_0=-0.01$, $r_V=622\,$Mpc/$h$, $D_V=7400\,$Mpc/$h$) for NILC.

\begin{figure}
    \centering
    \includegraphics[width=0.49\columnwidth]{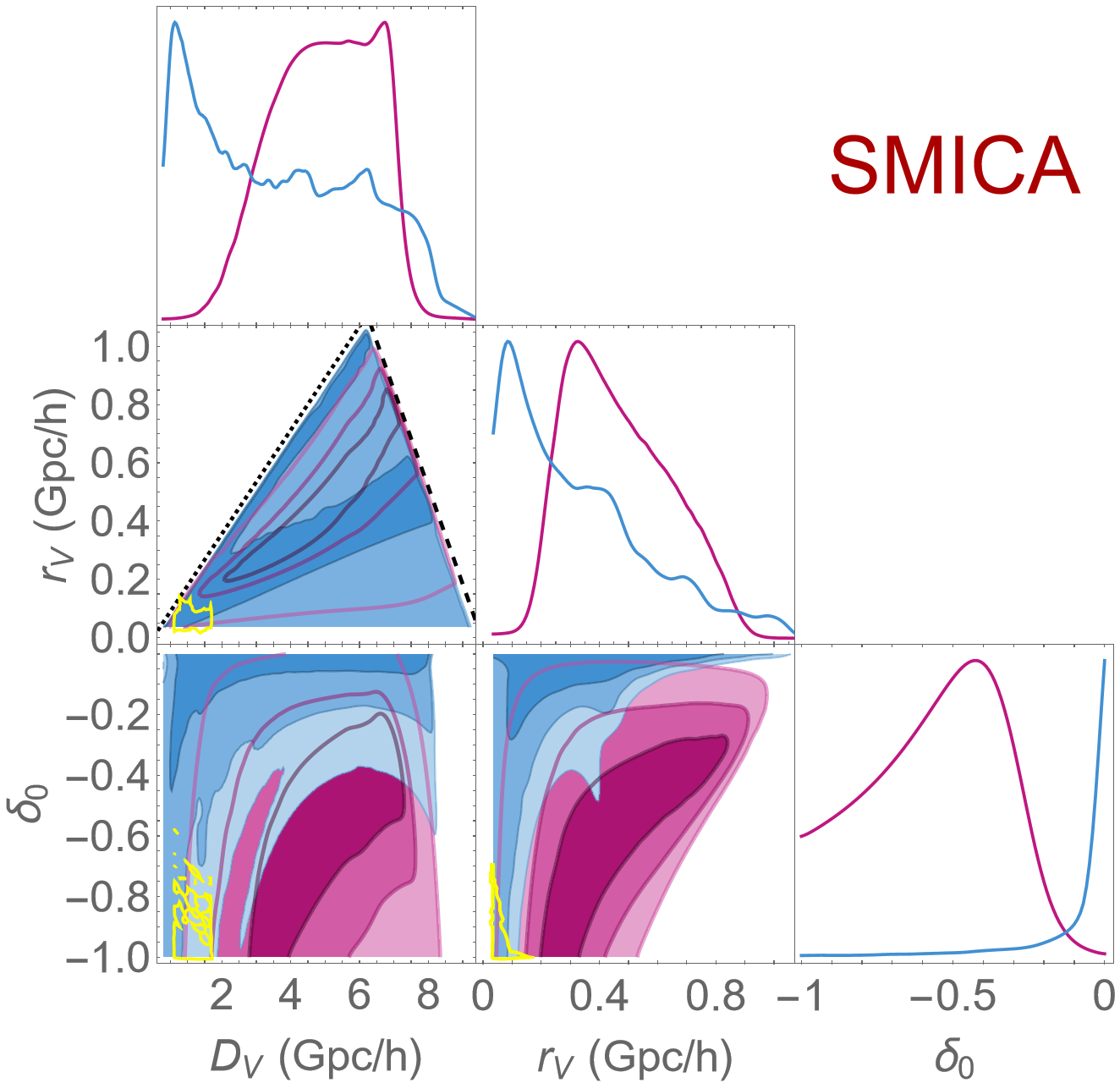}
    \includegraphics[width=0.49\columnwidth]{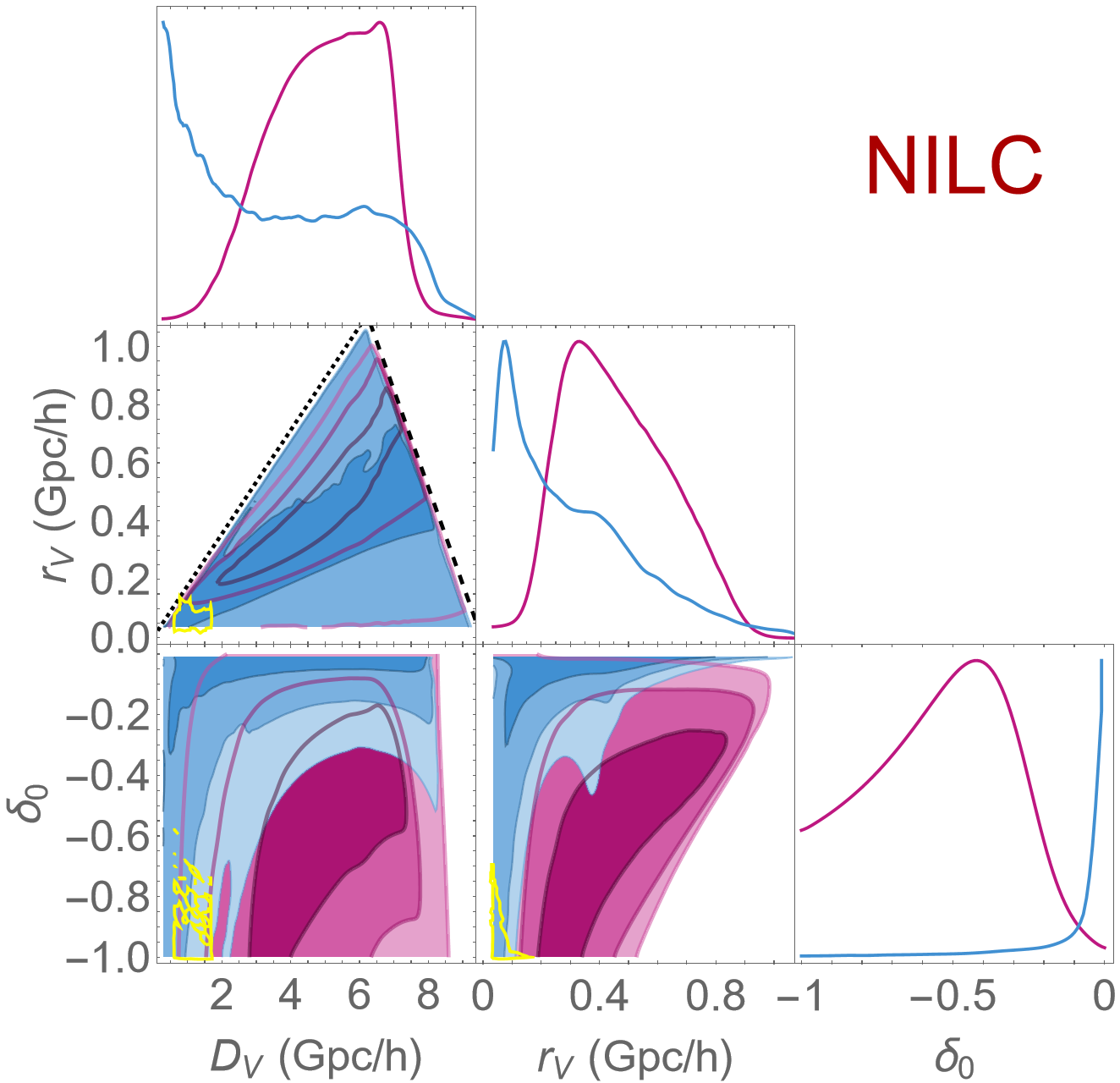}
    \caption{Triangular plots of the results for temperature (magenta) and lensing (blue) for comoving distance to the void $D_V$, comoving void radius $r_V$ and central density contrast $\delta_0$. 1, 2 and $3\sigma$ contours are shown for SMICA (left) and NILC (right).  The dashed (dotted) lines are our top-hat prior boundaries for the void position $D_V + 3r_V \leq D_{\rm SLS}$  (void angular size $\theta_V\leq 10.1^\circ$). Yellow contours depict the 95\% densest region in parameter space of real observed voids in the SDSS BOSS DR12 catalog [see text].
    \label{fig:triangular}
    }
\end{figure}

\begin{figure}
    \centering
    \includegraphics[width=0.49\columnwidth]{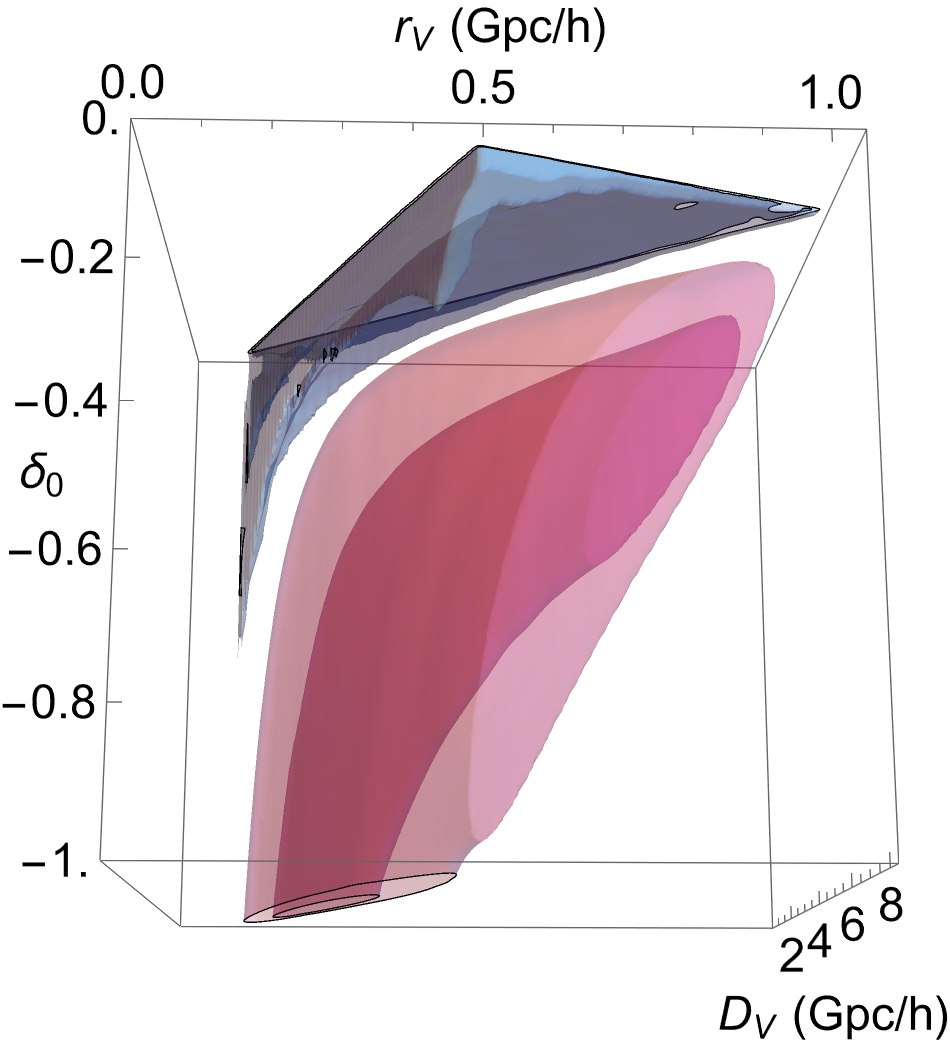}
    \includegraphics[width=0.49\columnwidth]{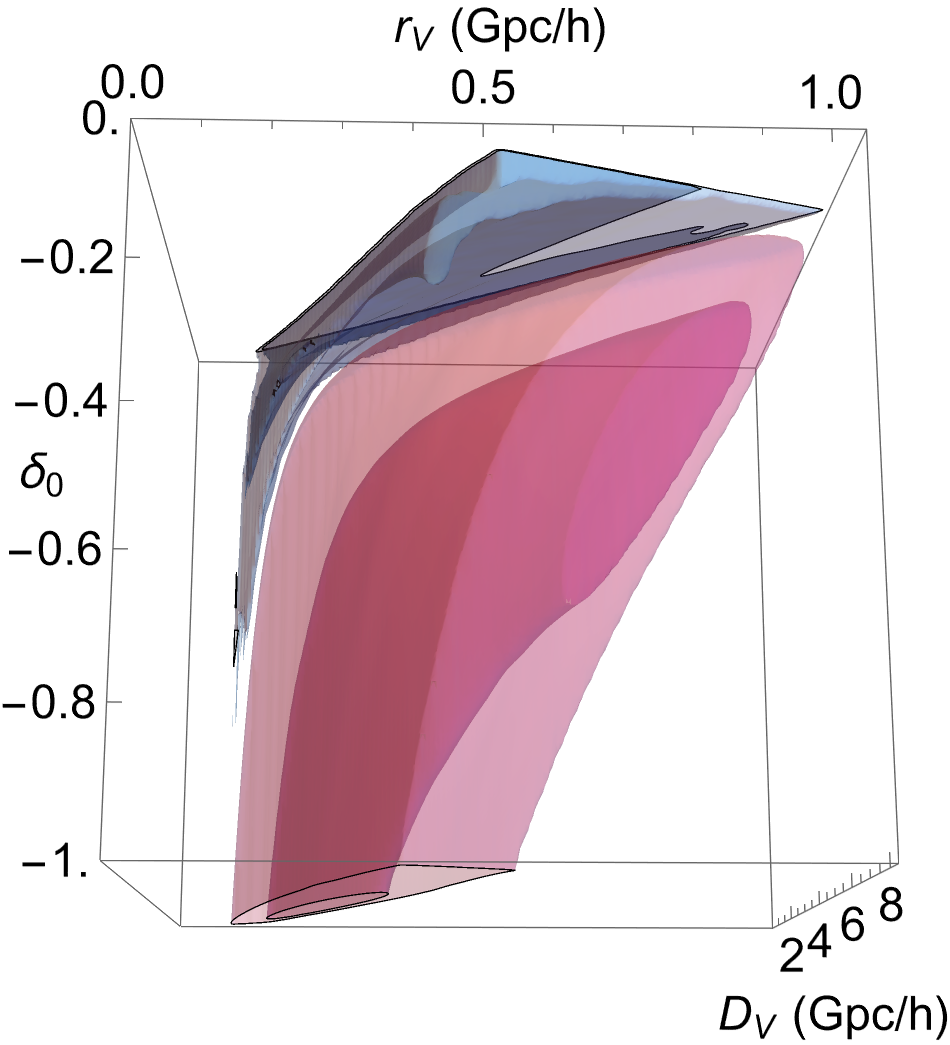}
    \includegraphics[width=0.49\columnwidth]{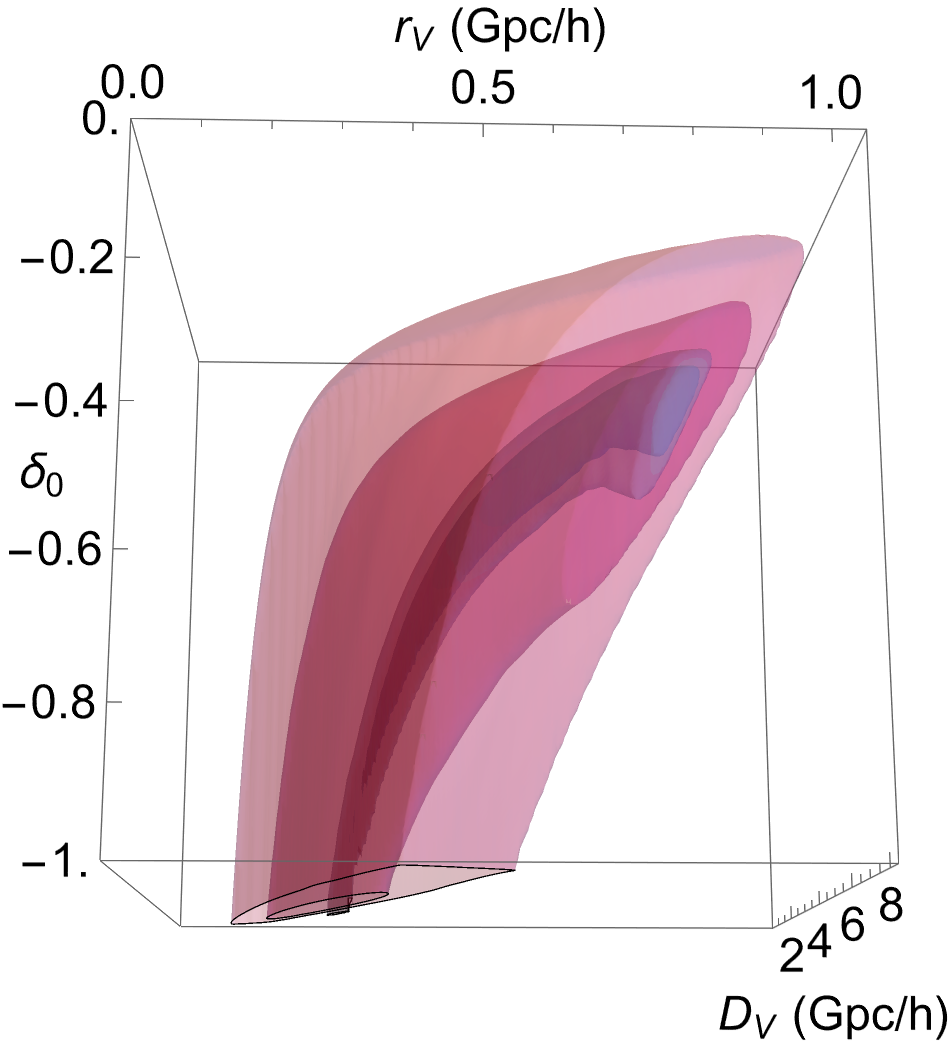}
    \includegraphics[width=0.49\columnwidth]{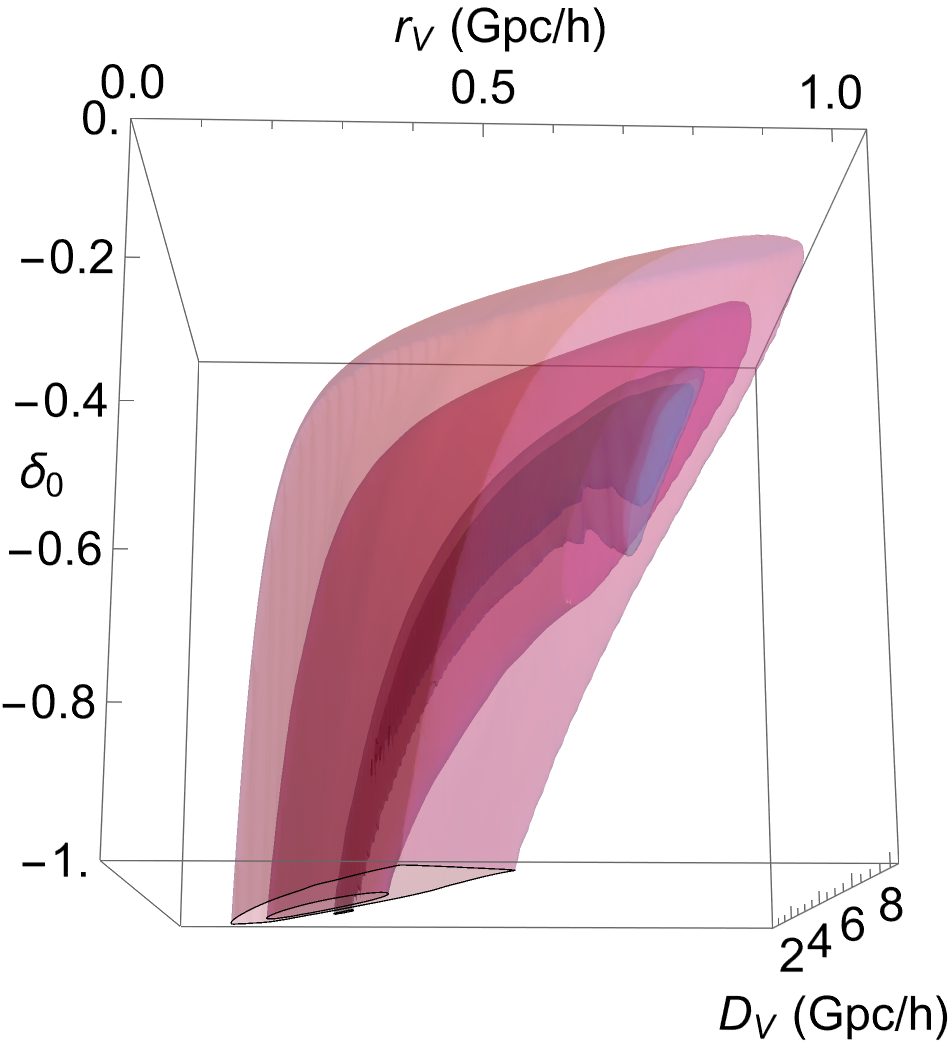}
    \caption{\emph{Top:} similar to Figure~\ref{fig:triangular} but for the 1 and $2\sigma$ 3D contours for SMICA (left) and NILC (right). The discrepancy between the temperature and lensing results are very clear as there is no overlap of the $2\sigma$ regions. \emph{Bottom:} replacing lensing data with simulations assuming the best-fit temperature parameters and either our standard Gaussian profile (left) or a lensing profile fitting DES data on voids (right).  The results are now in complete agreement, independent of the details of the profile model.
    \label{fig:3D-data}
    }
\end{figure}

Here, and in what follows, we will denote simply by 1, 2 and 3$\sigma$ the regions in which 68.3, 95.4 and 99.73\% of the posteriors are enclosed, respectively, even though they are highly non-Gaussian.
Figure \ref{fig:triangular} shows the 1, 2 and 3$\sigma$ contour plots for $\delta_0$, $r_V$ and $D_V$, for both temperature and  lensing. Temperature data strongly prefers a void located at $D_V \ge 2500$ Mpc/$h$, which corresponds to $z \ge 1$, a region poorly probed by current galaxy catalogs. The reason temperature data disfavours a void at low redshifts is that in order to fit the angular scale of the CS a low $r_V$ is needed, which leads to insufficient ISW decrease in temperature. Note also that this preference for high $z$ implies that the RS effect, which leads to a hot ring, can have a non-negligible role in the temperature profile. Lensing results instead are perfectly compatible with the absence of any significant voids at any redshifts. In Figure~\ref{fig:3D-data} we show similar plots, but for 1 and 2$\sigma$ contours in 3D. In these figures one can see more clearly that there is no overlap in 3D of the 2$\sigma$ regions, which demonstrates a tension between temperature and lensing. We also tested how much the results depend on the void profile by using the average DES void curvature profile mentioned previously. We used HEALPix Void to include a void with both profile choices, but found no noticeable difference, as can be seen in Figure~\ref{fig:3D-data}.

It is interesting to check if actually observed voids are in the detectable lensing region of the parameter space. We thus examined the SDSS BOSS DR12 void catalog~\cite{Mao:2016onb}. This catalog provides the central density contrast, radius and redshift for 1228 voids using the SDSS Data Release 12 galaxy catalog. As it is obtained using the ZOBOV void finding algorithm, based on Voronoi tessellation, we need to convert their effective radius $r_{\rm eff}$ to an equivalent $r_V$.\footnote{The effective radius is the radius of the sphere with the same volume of the Voronoi tesselation fitted region, which comprehends only the underdensity without any compensated part.  To convert it to our $r_V$ definition we considered only the underdense part of our profile, which extends up to $r=1.73r_V$. This implies that $r_V=r_{\rm eff}/1.73$.}
This results in an effective $r_V$ range $10-260$ Mpc/$h$. We depict this catalog with yellow contours in Figure \ref{fig:triangular}, which mark the 95\% densest region in this parameter space of this data. We remark that this catalog is known to have low completeness for $\delta_0 > -0.8$ and is limited to $z<0.7$, such that these yellow contours can only serve as a qualitative illustration. With these limitations in mind, we nonetheless find that the lensing contours seem consistent with what would be produced by a typical standard $\Lambda$CDM void in the cold spot direction.

In order to quantify how much the void model for the Cold Spot is disfavoured, we computed also the Bayes factor, which is the ratio of the Bayesian evidences. This has advantages over the Bayesian or Akaike information criteria as it makes less assumptions on properties of the distributions. More specifically, we computed the Bayes factor ${\mathcal B}_{01}$ between the two competing models: Model 0, pure $\Lambda$CDM with no free parameters, and Model 1, our void model with 3 parameters. For this purpose we considered temperature and lensing data together, which amounts to a multiplication of the likelihoods, assuming they are independent. We note that the $\Lambda$CDM likelihood ${\cal L}_0$ can be simply obtained as the particular case of the void model likelihood ${\cal L}_1$ when $\delta_0 = 0$. Care must be taken however to the fact that the top-hat prior  in model 1, ${\cal P}_1$, is not unity, so that to get ${\mathcal B}_{01}$ we must compute the following ratio:
\begin{equation}
    {\mathcal B}_{01} \,=\, \frac{{\cal L}_0}{\int {\cal L}_1(r_V, D_V, \delta_0) {\cal P}_1(r_V, D_V, \delta_0) \dd r_V \dd D_V \dd \delta_0}\,.
\end{equation}
In our case, as stated before we assume a top-hat prior ${\cal P}_1$. For SMICA (NILC) we found that $\ln {\mathcal B}_{01} =  2.6$ (3.0), disfavoring the void model. If we moreover consider the model space to consist of only these two models, we can assign probabilities to the Bayes factors~\citep[see][]{Trotta:2008qt}. Assuming equal model priors for both cases (no a priori preference for either) we found that the void model is favored with a probability of only 7.4\% (5.0\%) for SMICA (NILC). In other words, the $\Lambda$CDM is favored with odds 1 : 13 (1 : 20). If one repeats the same calculation now using only temperature data, one instead finds that the void model is favoured with odds 56 : 1 (21 : 1) for SMICA (NILC), which is expected as the void interpretation was made \emph{a posteriori} to fit the temperature data. Lensing therefore flips completely these odds against the void interpretation.

We remark that our likelihood behind both Figures~\ref{fig:triangular} and \ref{fig:3D-data} were computed assuming for simplicity a linear relation between $k_0$ and $\delta_0$, Eq.~\eqref{k0}. There is certainly some corrections to this at high values of $|\delta_0|$, which means that our prior choice $\delta_0 \ge -1$ corresponds in practice to slightly less negative densities in the past light cone. In other words, the contours at large negative values of $\delta_0$ would be mapped into slightly different values of $\delta_0$ if the full non-linear relation between $k_0$ and $\delta_0$ was used.  Nevertheless this should have only marginal impact on the discrepancy between lensing and temperature since our real data lensing contours are predominantly in the low density contrast regime, as is clearly shown in Figure~\ref{fig:3D-data}.

%%%%%%%%%%%%%%%%%%%%%%%%%%%%%%%%%%%%%%%%
%%%%%%%%%%%%%%%%%%%%%%%%%%%%%%%%%%%%%%%%
\section{Conclusions}\label{sec:conclusions}

We have searched for a signal in CMB lensing, assuming that the CMB Cold Spot is due to a large void located along the line of sight, with the only restriction that it should not reach the surface of last scattering itself nor the observer. If such a void were responsible for the CS temperature decrement due to the ISW and RS effects, it should produce a very significant lensing signature for a very broad range of possible redshifts. The successful tests using simulated supervoids show that we can recover the correct parameters through the lensing signal analysis. We nevertheless found no significant lensing signal in the real data. The absence of lensing is strong enough to flip the Bayes factor analysis from favouring to disfavouring the model with a void compared to $\Lambda$CDM with moderate evidence, even though such a model was already artificially favoured by the \emph{a posteriori} choice to fit the CS temperature profile.

By analyzing the data in both the SMICA and NILC maps from Planck 2018 CMB data we can confirm that our results are very similar in both cases, which makes our analysis of temperature and lensing signals robust against possible systematic effects. It also illustrates the robustness of our approach of removing biases.

Interestingly, one possible reason for which the void interpretation had not been ruled out before is that incidentally the angular size of the cold spot is precisely the one which is harderst to constrain with ISW and lensing, as first shown in~\cite{Zibin:2014rfa}. Smaller sizes would produce more lensing, larger sizes more ISW. We also note that as shown in that work, if a supervoid is located at redshifts around reionization, the kinetic Sunyaev Zel'dovich effect can also be used to constrain it. We tested limiting our parameter space to exclude such high redshifts, but our results do not change much.

Other interpretations for the Cold Spot remain open: (i) a random fluctuation in a Gaussian realization in the standard $\Lambda$CDM model, (ii) an anomalous structure located on the last scattering surface, (iii) more exotic possibilities such as a texture, (iv) the possibility that the ISW effect from late time voids (e.g.~from the \emph{Eridanus} void) could be magnified by a factor $A_{\rm ISW}\gtrsim 5$ of unknown origin, that would not affect lensing. Note that in interpretation (i) a correlated signal in polarization is expected~\cite{Vielva:2010vn}, but due to the low signal-to-noise in the Planck polarization data this could not be put to test with high significance~\cite{Planck:2019evm}.

In the two decades since its first identification, the Cold Spot has encouraged a number of follow-up galaxy surveys in its area. Although a couple of voids have been detected~\cite{Kovacs:2021wnc} in the intervening large-scale structure, galaxy surveys are still limited in their redshift coverage. By leveraging the lensing information on the CMB data we were able to search for a large void in the complete redshift range separating us from the CMB. The absence of a clear lensing signature now puts the supervoid interpretation to rest.

%%%%%%%%%%%%%%%%%%%%%%%%%%%%%%%%%%%%%%%%
%%%%%%%%%%%%%%%%%%%%%%%%%%%%%%%%%%%%%%%%

%%%%%%%%%%%%%%%%%%%%%%%%%%%%
%% END OF SECTION
%%%%%%%%%%%%%%%%%%%%%%%%%%%

\emph{\textbf{Acknowledgments.}}
We thank András Kovács for providing access to the DES void catalog and convergence measurements and Seshadri Nadathur, Krishna Naidoo and James Zibin for interesting comments. SO is supported by Brazilian research agencies CNPq and FAPERJ. PF is supported by the Brazilian research agency CAPES. The work of A.N. is supported by the grants PID2019-108122GB-C32 from the Spanish Ministry of Science and Innovation, Unit of Excellence Mar\'ia de Maeztu 2020-2023 of ICCUB (CEX2019-000918-M) and AGAUR2017-SGR-754. A.N. is grateful to the IFPU (SISSA, Trieste) and to the Physics Department of the University of Florence for the hospitality during the course of this work.
A.N. acknowledges the Spanish Ministerio de Universidades, the Next Generation EU plan from the European Union and the Recovery, Transformation and Resilience Plan, for financing his scientific visit at GGI, Florence.
MQ is supported by the Brazilian research agencies CNPq, FAPERJ and CAPES. This study was financed in part by the Coordenação de Aperfeiçoamento de Pessoal de Nível Superior - Brasil (CAPES) - Finance Code 001. We acknowledge support from the CAPES-DAAD bilateral project  ``Data Analysis and Model Testing in the Era of Precision Cosmology''. The authors acknowledge the National Laboratory for Scientific Computing (LNCC/MCTI, Brazil) for providing HPC resources of the SDumont supercomputer (\url{http://sdumont.lncc.br}), which have contributed to the research results reported within this paper.

\appendix\label{sec:app}

%%%%%%%%%%%%%%%%%%%%%%%%%%%%%%%%%%%%%%%%
%%%%%%%%%%%%%%%%%%%%%%%%%%%%%%%%%%%%%%%%
\section{Pipeline and lensing estimator details}\label{appendix:pipeline}
%%%%%%%%%%%%%%%%%%%%%%%%%%%%%%%%%%%%%%%%
%%%%%%%%%%%%%%%%%%%%%%%%%%%%%%%%%%%%%%%%

%%%%%%%%%%%%%%%%%%%%%%%%%%%%%%%%%%%%%%%%
\subsection{Pipeline flowchart}
%%%%%%%%%%%%%%%%%%%%%%%%%%%%%%%%%%%%%%%%

\begin{figure}[t]
    \centering
    \includegraphics[width=0.76\linewidth]{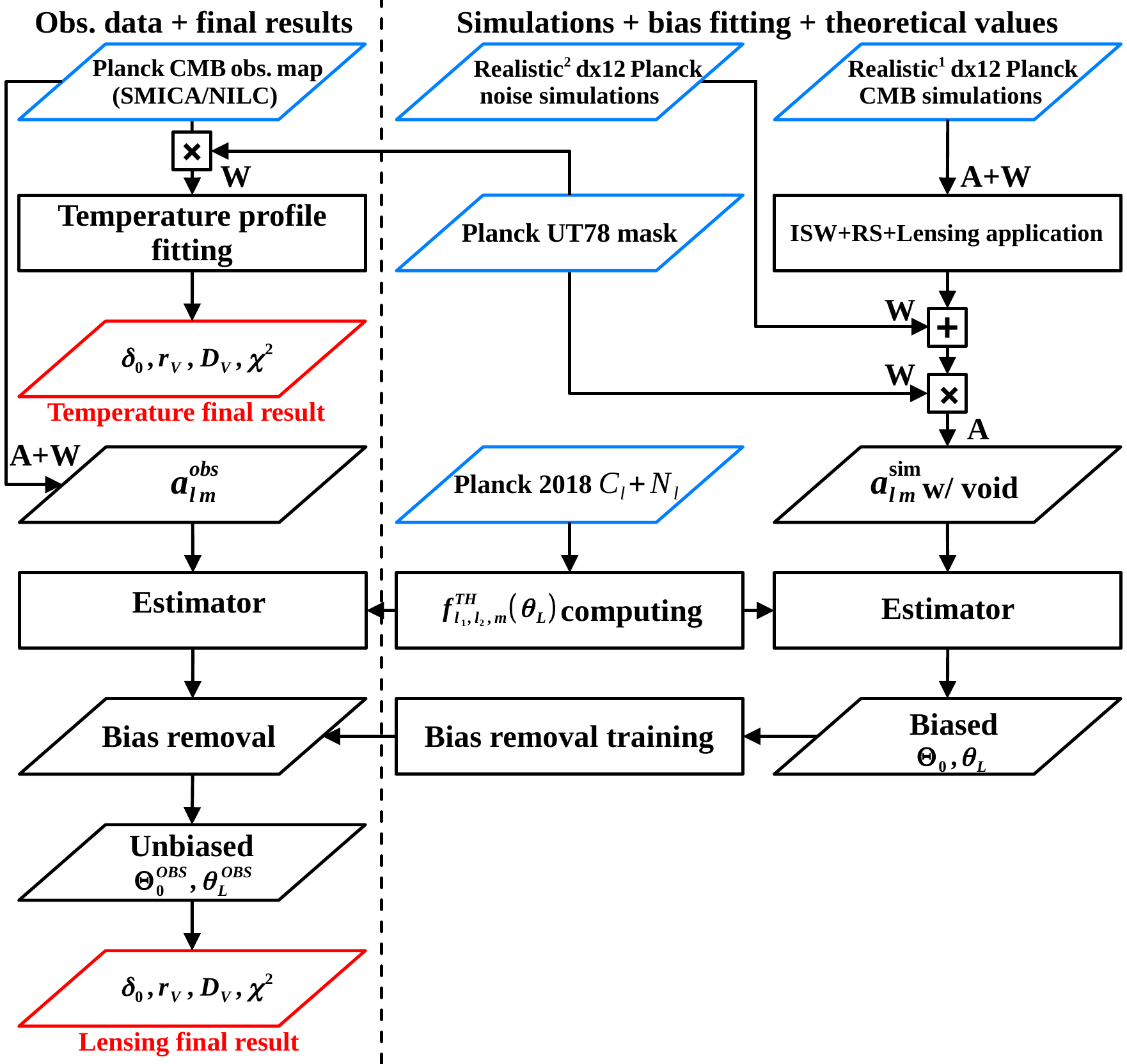}
    \caption{Our main pipeline. Blue boxes are input files from Planck, red are the final outputs. \emph{Left column:} path from observational data to the final results for lensing and temperature. \emph{Middle and right columns:} path from Planck simulations to the mock simulations with ISW, RS and lensing applied,  $f^{\rm TH}_{\ell_1,\ell_2,m}$ computation and bias removal. W means a Wigner rotation (pointing the void to the north pole) and A the $a_{\ell m}$ calculation with HEALPix's Anafast.}\label{fig:pipeline}
\end{figure}

As explained in Section \ref{sec:Methodology}, in order to estimate the temperature and lensing effects due to a void we employed a large set of realistic simulations, with the purpose of training the bias removal method in the lensing estimator. We also converted the two lensing parameters into the three physical parameters $r_V$, $D_V$ and $\delta_0$, which allows comparison of the results in the same parameter space as temperature. Due the complexity of the whole process, we have summarized it in  Figure~\ref{fig:pipeline}. For further details on the block ``ISW+RS+Lensing application'' see  section \ref{appendix:sims}, for ``Estimator'' see  section~\ref{sec:lens estimator}, and finally for the description of the lensing ``Bias removal'' process see section~\ref{appendix:bias_plot}. Using this procedure we generated $\sim 53000$ simulations over a grid of $1550$ different combinations of $\Theta_0$ and $\theta_V$, plus 10000 simulations with random values of $\Theta_0$ and $\theta_V$. We also created 6000 null simulations (without void) following the same pipeline but skipping the ``ISW+RS+Lensing application'' step. We did it for both SMICA and NILC maps.

%%%%%%%%%%%%%%%%%%%%%%%%%%%%%%%%%%%%%%%%
\subsection{Simulations with lensing, ISW and Rees-Sciama}\label{appendix:sims}
%%%%%%%%%%%%%%%%%%%%%%%%%%%%%%%%%%%%%%%%

\begin{figure}[t]
    \centering
    \includegraphics[width=0.447\linewidth,trim={0 -0.2cm 0 0},clip]{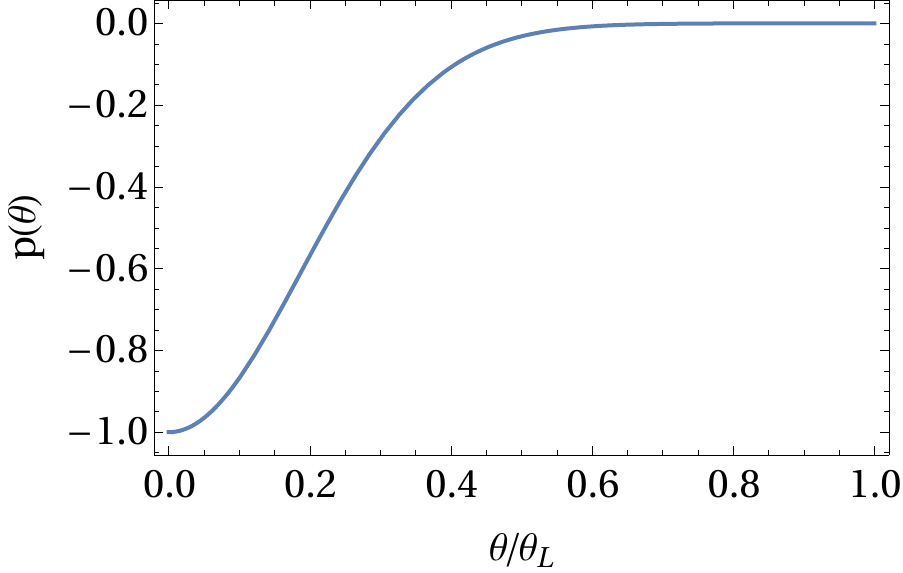}
    \includegraphics[width=0.487\linewidth,trim={0 -0.2cm 0 0},clip]{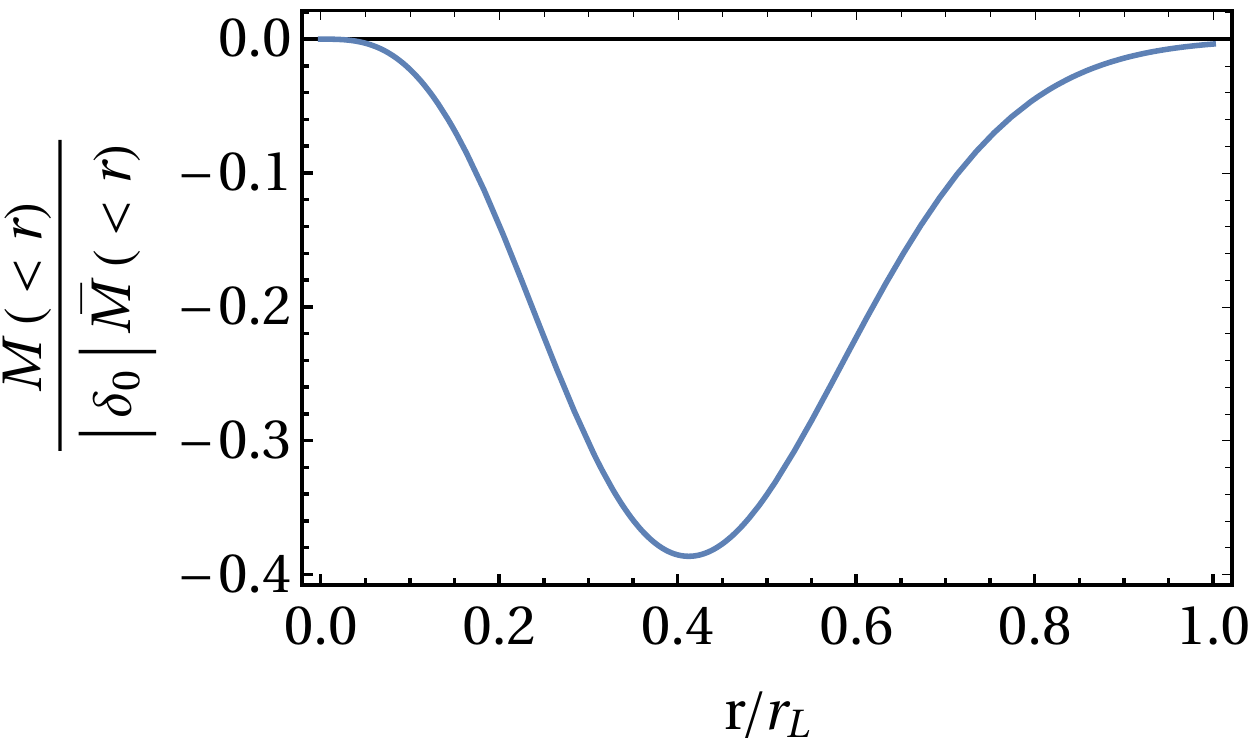}
    \caption{\textit{Left:} The lensing profile $p(\theta)$ as a function of $\theta/\theta_L$. \textit{Right:} The accumulated mass $M(<r) = \int^r_0 4\pi \rho \, r'^2 \dd r'$ as function of the radius, divided by the average mass $\bar{M}$ and normalized to $|\delta_0|$. At $4.2r_V$ ($\theta=\theta_L$) the void is already 99\% compensated.
    } \label{fig:P_theta}
\end{figure}

Here we summarize the process of applying ISW, RS and lensing on simulations. For simplicity we first apply all the effects on the $\hat{z}$ direction and then rotate the $a_{\ell m}$'s to the desired directions using Wigner matrices, using HEALPix. As the void profile is spherical, all effects will depend only on $\theta$.
For temperature, the ISW and RS effects are introduced summing the Planck CMB E2E simulation maps with the equations \eqref{ISWintheta} and \eqref{RSintheta}
\begin{eqnarray}
    \frac{\Delta T^{\rm ISW+RS}}{T_0}(\theta,\phi) = \frac{\Delta T^{\rm P}}{T_0}(\theta,\phi) + \frac{\Delta T^{\rm ISW}}{T_0}(\theta) + \frac{\Delta T^{\rm RS}}{T_0}(\theta) \,.
\end{eqnarray}
The next step is to apply the lensing effect due to the void. This is done in the conversion between spherical harmonics and pixel space, using a modification of the HEALPix Boost code~\cite{Notari:2013iva,Ferreira:2021omv},\footnote{\url{https://github.com/mquartin/healpix-boost}} that we called HEALPix Void. To proceed we need the lensing angle $\alpha$ which comes from the lensing potential Eq.~\eqref{lensing_potential} as
\begin{equation}\label{alpha}
    \alpha(\theta) = \frac{\partial \Theta(\theta)}{\partial \theta} = \Theta_0 \frac{\partial p(\theta)}{\partial\theta} \, .
\end{equation}
We fit $p(\theta)$ with a $20^{\rm th}$ order polynomial in the variable $\theta/\theta_L$, to reach high precision, with the constraint $\theta<\theta_L$. We show the plot of $p(\theta)$ in Figure \ref{fig:P_theta}. We also show in this Figure the accumulated void mass as a function of radius, which justifies our choice of $r_L$ (to wit, $r_L = 4.2 r_V$) and thus of $\theta_L$.

HEALPix Void then changes the angle $\theta$ on the $Y_{\ell \, m}$ functions by adding $\alpha$:
\begin{equation}\label{Ylm}
    Y_{\ell \, m}(\theta, \phi) \rightarrow Y_{\ell \, m}\big(\theta+\alpha(\theta),\phi\big)  \, .
\end{equation}
 The final map in pixel space is obtained as
\begin{eqnarray}
    \frac{{\Delta T}^{\rm ISW+RS+L}}{T_0}(\theta)=\sum_{\ell,m} a^{\rm ISW+RS}_{\ell \, m}Y_{\ell \, m}\big(\theta+\alpha(\theta),\phi\big).
\end{eqnarray}

%%%%%%%%%%%%%%%%%%%%%%%%%%%%%%%%%%%%%%%%
\subsection{Lensing bias removal and $\chi^2$ grid conversion to physical parameters}\label{appendix:bias_plot}
%%%%%%%%%%%%%%%%%%%%%%%%%%%%%%%%%%%%%%%%
We estimated the $\Theta_0$ and $\theta_V$ parameters from the simulations obtained following our pipeline 
and used both to train two bias removal methods, the Random Forest Regression (RFR), from the scikit-learn library~\citep{Pedregosa:2011ork}, and a $3^{rd}$ order polynomial fit on the grid of estimated values for $\Theta_0$ and $\theta_V$. RFR is a machine learning method based on the weighted average of decision trees outputs, where each decision tree is a discrete estimator that can recover a value provided during the training process based on a sequence of conditional operations. These trees are trained using a bootstrap sample of the training sample to minimize the error of the predicted regression using the inputs and expected (fiducial) outputs \cite{breiman2001random}. We tested the trained bias using the 10000 simulations with random values of $\Theta_0$ and $\theta_V$, and 6000 null simulations (without void) to avoid over-fitting, verifying that both methods work reasonable well. We got significantly better performance for the RFR method using a forest of 1024 trees compared to the $3^{\rm rd}$ order polynomial fit, and thus we chose it as our main method. We trained one forest for each parameter pair $\Theta_0,\, \theta_V$; the inputs were the biased $\Theta_0^{\rm {sims,b}}$ and $\theta_V^{\rm {sims,b}}$ (b for biased) and the outputs were the estimate for the fiducial values considered in the simulations, $\Theta_0$ or $\theta_V$. Therefore each forest takes two inputs ($\Theta_0^{\rm {sims,b}}$, $\theta_V^{\rm {sims,b}}$) and gives one output ($\Theta_0^{\rm {sims,u}}$ or $\theta_V^{\rm {sims,u}}$), where the u superscript stands for unbiased.

The bias was tested for the estimator in several maximum multipole combinations and the conclusion is that the optimal values to use for Planck data are $\ell_{\rm max}=1600$ and maximum non-diagonal distance $\Delta \ell \equiv \ell_2 - \ell_1 = 40$. They provide a good combination of S/N, residual bias and computational time. In fact, the computational cost is significant, and the full pipeline took $\sim45000$ CPU hours using a $3^{\rm rd}$ gen AMD EPYC mini-cluster. We found that the same trained bias removal procedure works equally well with both our Gaussian profile and the DES lensing profile.

\begin{figure}
    \centering
    \includegraphics[width=0.91\linewidth]{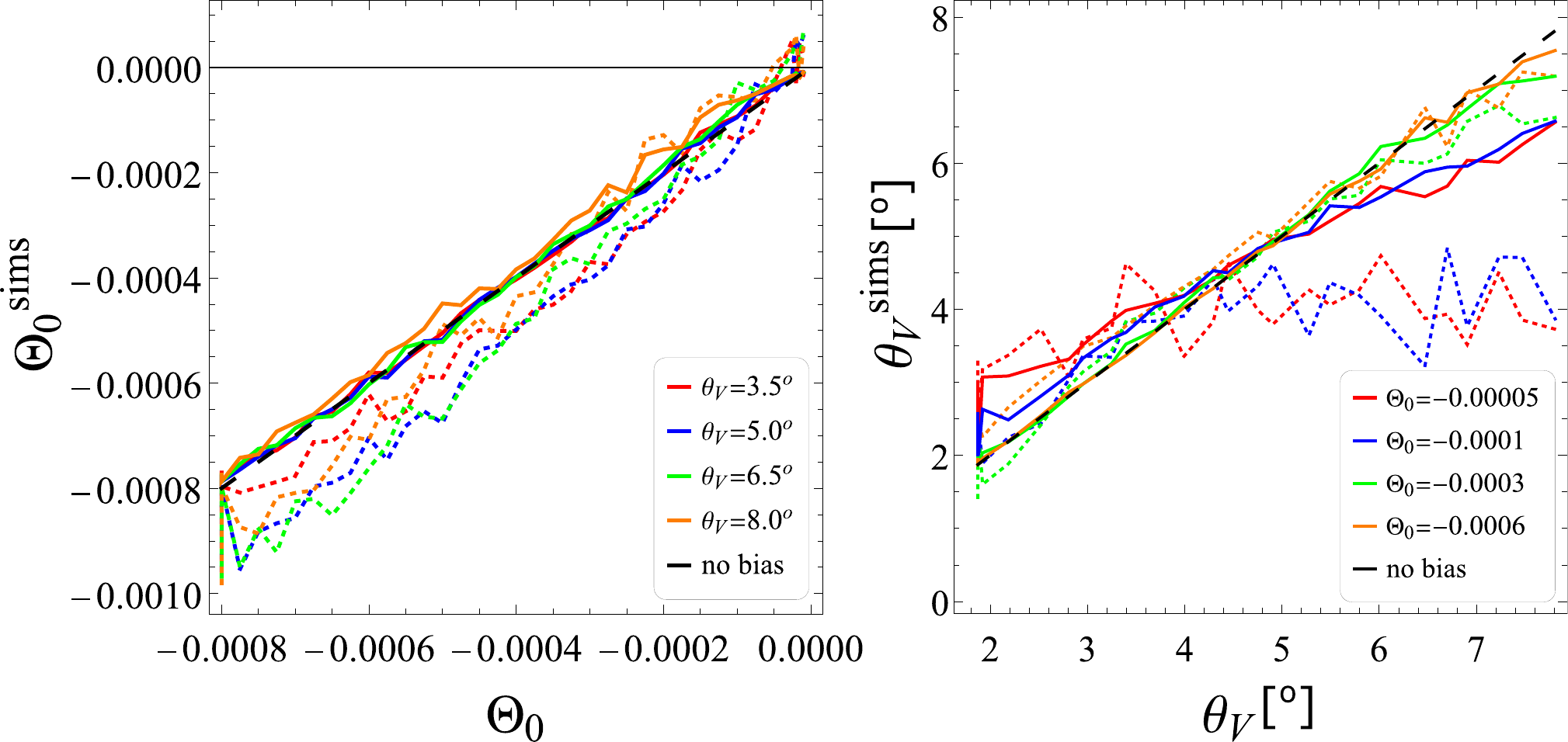}
    \caption{Tests of the lensing estimator using simulations. In the horizontal axis we show the fiducial values (expected values) and in the vertical axis the recovered values from simulations. Dotted (continuous) lines represents the naïve/biased (bias removed) results. The dashed line is the ideal case without any bias.  \emph{Left:} recovered values of $\Theta_0$ for 4 values of $\theta_V$. \emph{Right:} same for $\theta_V$ for 4 values of $\Theta_0$.
    }\label{fig:bias_plot}
\end{figure}

In Figure \ref{fig:bias_plot} we show the effectiveness of the bias removal for some cases with either $\Theta_0$ or $\theta_V$ fixed. We concluded that the bias removal method is satisfactory. However, if the void is very shallow (very small $\Theta_0$), while the residual bias on $\Theta_0$ is still small, the parameter $\theta_V$ cannot be measured well. This is because the size becomes almost irrelevant. This implies that $\theta_V$ will have some residual bias in these cases, as can be seen on the right panel.

In order to compare our lensing and temperature results we first mapped the biased $\chi^2(\Theta_0^{\rm {sims,b}}, \theta_V^{\rm {sims,b}})$ into the unbiased $\chi^2(\Theta_0^{\rm {sims,u}}, \theta_V^{\rm {sims,u}})$ by changing the coordinates of each computed $\chi^2$ value to the unbiased position. We then perform a parameter transformation of the unbiased $\chi^2(\Theta_{0}^{\rm {sims,u}}, \theta_V^{\rm {sims,u}})$ 2D grid to the physical 3D grid $\{\delta_0,\,r_V,\,D_V\}$. This was achieved by computing the expected $\Theta_0$ and $\theta_V$ for each $r_V$, $D_V$ and $\delta_0$ combination, using equation~\eqref{lensing_potential} and considering $\theta_V=\arctan{(r_V/D_V)}$. We employed the grid $r_V  \in [35,\, 1770] \, \mathrm{Mpc}/h$ in steps of 14 Mpc/$h$, $D_V \in [290,\,9570] \,\mathrm{Mpc}/h$ in steps of 31 Mpc/$h$ and $-1\leq \delta_0\leq 0 $ in steps of 0.01. The same grid ranges were used in both temperature and lensing estimators.

\subsection{Optimizing the number of modes considered}%\label{appendix:lensing_estimator}

\begin{figure}[t]
    \centering
    \includegraphics[width=0.61\linewidth]{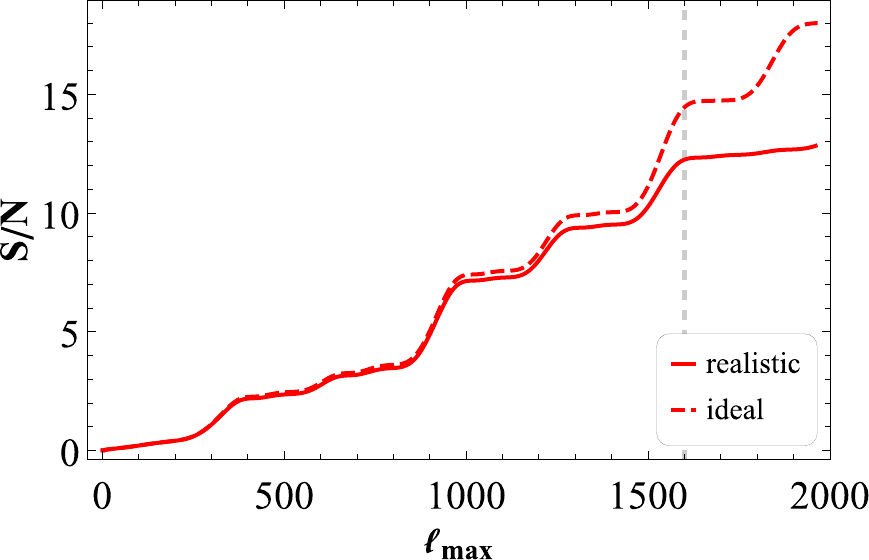}
    \caption{The cumulative signal-to-noise ratio in non-diagonal lensing correlations, computed assuming the best-fit temperature profile parameters from the CS, for $|\ell_1-\ell_2|\leq 40$.  Solid (dashed) curves represent the realistic (ideal) cases, where realistic means the effect of mask and noise over the $C_{\ell}$ were considered.    The vertical gray dashed line indicates $\ell_{{\rm max}}=1600$, the maximum multipole considered in the analysis.} \label{fig:StoN}
\end{figure}

The lensing $\chi^2$, Eq.~\eqref{chisq_lensing} which includes the change $C_\ell \rightarrow \Tilde{C}_\ell + \Tilde{N_\ell}$, is calculated over a grid with the ranges $-0.0011\leq \Theta_0 \leq 0$ and $\theta_V \leq 14.1^\circ$ (these ranges are before bias removal). The elements of the numerator of Eq.~\eqref{chisq_lensing}, to wit $f^{\rm OBS}_{\ell_1,\ell_2,m}$ and $f^{\rm TH}_{\ell_1,\ell_2,m}$, are matrices with dimensions $[\ell_{\rm max}(\ell_{\rm max}+3)/2+1] \times \Delta\ell$. As discussed in the main text, we set the maximum non-diagonal correlation length in multipole space, $\Delta \ell$, to $40$, and $\ell_{\rm max}=1600$. Higher $\ell_{\rm max}$ or $\Delta \ell$ increase only marginally the S/N. Figure~\ref{fig:StoN} uses Eq.~\eqref{StoN} with $C_\ell \rightarrow \Tilde{C}_\ell + \Tilde{N_\ell}$ to depict the dependence on $\ell_{\rm max}$. A larger bias at higher $\ell_{\rm max}$ is expected as the noise quickly dominates the Planck data.

\subsection{Rees-Sciama significance}\label{app:RS}
\begin{figure}[t]
\centering
    \includegraphics[width=0.61\linewidth]{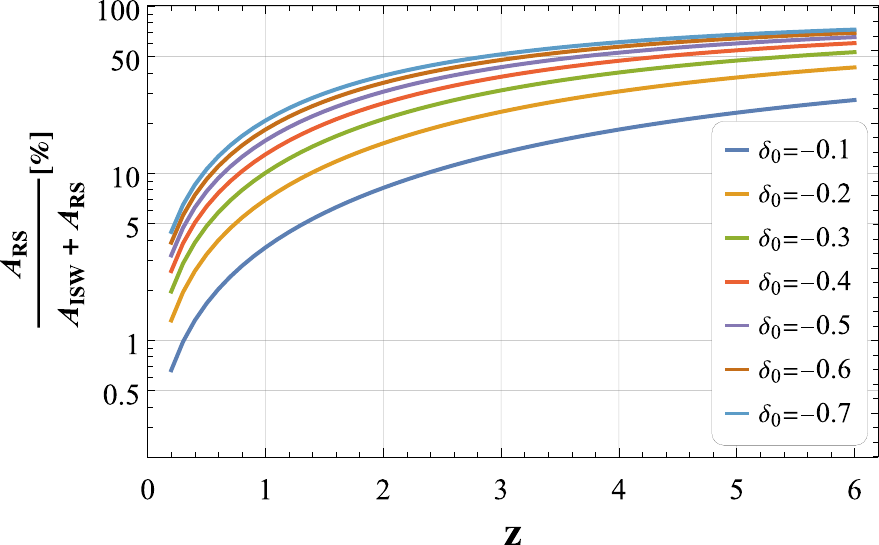}
    \caption{The ratio of the temperature fluctuation due to RS at the center, $A_{\rm RS}$, over the sum of the amplitude at the center due to ISW, $A_{\rm ISW}$, plus $A_{\rm RS}$,  as a function of redshift for given values of $\delta_0$. We keep $\theta_V$ fixed at $6.3^\circ$, which corresponds to a void radius $r_V$ in a range between 40 ($z=0.1$) and 830 Mpc/$h$ ($z=6$).}\label{deltaT_amplitudes}
\end{figure}

Figure \ref{deltaT_amplitudes} shows the ratio of the RS effect over the amplitude of  ISW + RS at the center, as a function of redshift. This illustrates that the RS contribution is significant if the void is located at $z\gtrsim 2$, which as discussed in the main text, incidentally is the preferred region from the fit of the temperature data.

\bibliographystyle{JHEP2015}
\bibliography{cmb}

\end{document}